\documentclass[12pt]{article}
\usepackage{epsfig,amsmath,amssymb,subfigure,enumerate}
\usepackage{caption2}

   {\end{list}}%

\makeatletter
\renewcommand{\section}{\setcounter{equation}{0}\@startsection
  {section}%
  {1}%
  {0pt}%
  {-1\baselineskip}%
  {0.4\baselineskip}%
  {\bfseries\large}}%
\renewcommand{\subsection}{\@startsection
  {subsection}%
  {2}%
  {0pt}%
  {-0.75\baselineskip}%
  {0.2\baselineskip}%
  {\bfseries}}%
\renewcommand{\subsubsection}{\@startsection
  {subsubsection}%
  {3}%
  {0pt}%
  {-0.5\baselineskip}%
  {0.1\baselineskip}%
  {\sc}}%
\makeatother

\renewcommand{\theequation}{\thesection.\arabic{equation}}

\def\a{\alpha}
\def\b{\beta}

\def\m{\mu}
\def\n{\nu}
\def\r{\rho}
\def\s{\sigma}

\def\Dcal{{\mathfrak{D}}}
\def\Dirac{{D\mkern-12mu/}}

\def\pslash{{p\mkern-8mu/}{\!}}

\def\pslash  {{p\mkern-7mu/}}

\def\xslash  {{x\mkern-7mu/}}
\def\yslash  {{y\mkern-7mu/}}

\def\bp{\text{\tiny{BPST}}}
\def\id{{\rm{I}\!\rm{I}}}
 \def\tr{\text{Tr}}
 
\def\idp{\int\!\! \frac{d^4\!p}{(2\pi)^4}}

\def\idx{\int\!\! d^4\!x}

\def\rig>{\right>}

\newcommand{\bea}{\begin{eqnarray}}
\newcommand{\eea}{\end{eqnarray}}
\newcommand{\beann}{\begin{eqnarray*}}
\newcommand{\eeann}{\end{eqnarray*}}
\newcommand{\ba}{\begin{array}}
\newcommand{\ea}{\end{array}}

\newcommand{\Tr}{\mathbf{Tr}}

\def\psib{\overline{\psi}}

\def\g5{\gamma_{5}}

\def\pslash  {{p\mkern-7mu/}}
\def\Dcal{{\mathfrak{D}}}

\def\idx{\int\! d^{4}\!x\,}
\def\idy{\int\! d^{4}\!y\,}

 \def\Dirac{{D\mkern-12mu/}\,}

 \def\pslash  {{p\mkern-7mu/}}

 \def\Dcal{{\mathfrak{D}}}
 
 \def\Db {{\partial}_{\beta}}
 \def\Dm {{\partial}_{\mu}}
 \def\Dn {{\partial}_{\nu}}

 \def\aa {a_{\alpha}}
 
 \def\am {a_{\mu}}

 \def\g {\gamma}

 \def\a {\alpha}
\def\b {\beta}
\def\r {\rho}
 \def\s {\sigma}
\def\Qc{{\mathcal{Q}}}
\def\Rc{{\mathcal{R}}}
\def\Sc{{\mathcal{S}}}

 \def\Tr{\text{Tr}}
\def\tr{\text{tr}}

\begin{document}
\begin{titlepage}
\rightline{FTI/UCM 100-2005}
\vglue 45pt

\begin{center}

{\Large \bf Noncommutative QCD, first-order-in-$\theta$-deformed instantons and 't Hooft
vertices.}\\
\vskip 1.2 true cm
{\rm C.P. Mart\'{\i}n}\footnote{E-mail: carmelo@elbereth.fis.ucm.es}
 and C. Tamarit\footnote{E-mail: ctamarit@fis.ucm.es}
\vskip 0.3 true cm
{\it Departamento de F\'{\i}sica Te\'orica I,
Facultad de Ciencias F\'{\i}sicas\\
Universidad Complutense de Madrid,
 28040 Madrid, Spain}\\
\vskip 0.75 true cm

\vskip 0.25 true cm

{\leftskip=50pt \rightskip=50pt \noindent For commutative Euclidean time, we
study the existence of field configurations that {\it a)} are
formal power series expansions in $h\theta^{\m\n}$, {\it b)} go to
ordinary (anti-)instantons as  $h\theta^{\m\n}\rightarrow 0$, and
{\it c)} render stationary the classical action of Euclidean
noncommutative $SU(3)$ Yang-Mills theory. We show  that the
noncommutative (anti-)self-duality equations have no solutions of
this type at any order in $h\theta^{\m\n}$. However, we obtain all the deformations --called
first-order-in-$\theta$-deformed instantons-- of the ordinary
instanton that, at first order in $h\theta^{\m\n}$, satisfy the
equations of motion of Euclidean noncommutative $SU(3)$ Yang-Mills
theory. We analyze the quantum effects that these field
configurations give rise to in noncommutative $SU(3)$ with one,
two and three nearly massless flavours and compute the
corresponding 't Hooft vertices, also, at first order in $h\theta^{\m\n}$.
Other issues analyzed in this paper are the existence at higher orders in
$h\theta^{\m\n}$ of topologically nontrivial solutions of the type
mentioned above and the classification of the classical vacua of
noncommutative $SU(N)$ Yang-Mills theory that are power series in
$h\theta^{\m\n}$.
\par }
\end{center}

\vspace{20pt}
\noindent
{\em PACS:} 

\vfill
\end{titlepage}


\setcounter{page}{2}
\section{Introduction}

Instantons play a major role in the understanding of the non-perturbative
properties of QCD. The solution of the $U(1)_A$ problem, the mass of
the $\eta'$ and the explanation of the spontaneous chiral symmetry breaking
in QCD furnish instances of issues where instantons are the leading actors
--see~\cite{Schafer:1996wv} and~\cite{Diakonov:2002fq} and references therein.
Two chief phenomena which are at the heart of instanton physics are following.
First, instantons interpolate in Euclidean time between two classical vacuum
states with winding numbers $n$ and $n+1$, respectively, thus yielding the
semi-classical contribution to the transition probability between these two
classical vacuum states. Secondly, in the presence of massless quarks, the
instanton transition leads to compulsory quark-anti-quark pair creation 
or, alternatively, turns a left handed quark into a right handed quark.

Instantons also occur in noncommutative $U(N)$ Yang-Mills theories, 
in spite of the fact that, even classically, they  are not invariant
under scale transformations. It all started with the construction
of instantons in noncommutative $U(1)$ theory by the authors of
ref.~\cite{Nekrasov:1998ss} --see also
ref.~\cite{Nekrasov:2000zz}.  These instantons have no counterpart in ordinary space. Then instantons in noncommutative
$U(2)$ theories were constructed~\cite{Furuuchi:1999kv,
Kim:2000ms, Kim:2001ai, Chu:2001cx}, and thus was obtained the
noncommutative counterpart of the celebrated BPST instanton
~\cite{Belavin:1975fg}. (Multi)-Instantons in noncommutative
$U(N)$ gauge theories have also been constructed in
refs.~\cite{Correa:2001wv, Ishikawa:2001ye, Lechtenfeld:2001ie, Parvizi:2002mg,
Pomeroy:2002ik, Sako:2002mq,
Horvath:2002bj} and~\cite{Ivanova:2003eq}. The physical effects of
the noncommutative $U(N)$ instantons have been analyzed in a
number of papers. We shall just mention that the zero modes of the
Dirac operator in a noncommutative instanton background have been
studied in ref.~\cite{Kim:2002qm} and that the quantum corrections
around such types of backgrounds have been worked out for $N=2$
supersymmetry in ref.~\cite{Hollowood:2001ng}.

Noncommutative QCD was constructed in
ref.~\cite{Calmet:2001na} as a part of the noncommutative standard
model --see also refs.~\cite{Jurco:2001rq}
and~\cite{Aschieri:2002mc} and see refs.~\cite{Khoze:2004zc,Abel:2005rh} for 
other approaches. In the generalization of ordinary QCD of 
ref.~\cite{Calmet:2001na}, 
the noncommutative gauge field does not take values  in the Lie
algebra of $SU(3)$, but rather in its enveloping algebra. Actually, 
the noncommutative fields are built from the ordinary fields with
the help of the Seiberg-Witten map~\cite{Jurco:2000ja}. Thus it
was circumvented what appears to be a shortcoming of the standard
framework of noncommutative gauge theories, namely that it can
only be applied to $U(N)$ groups. Indeed, in this standard
framework --see ref.~\cite{Szabo:2001kg} for a good introduction
to the subject-- the noncommutative gauge field unavoidably takes
values in the Lie algebra of $U(N)$, or direct products of such
groups~\cite{Terashima:2000xq}. Some
phenomenological~\cite{Melic:2005hb, Melic:2005su} and
theoretical~\cite{Martin:2002nr, Brandt:2003fx,  Martin:2005gt}
properties of noncommutative gauge theories with $SU(N)$ gauge
groups have been investigated so far, but a lot of work remains
to be done.In particular, the study of the existence of 
instantons and, would they exist, the phenomena they give rise to,
is, up to the best of our knowledge, a completely unexplored
territory. This is in sharp contrast with the case of noncommutative
$U(N)$ theories. Note that in the case at hand, the $SU(N)$ theory
is not included in the $U(N)$ case, since the noncommutative $SU(N)$
gauge field does not take values in the Lie algebra of $SU(N)$. Soliton 
solutions in theories with $U(1)$ symmetry defined by using the  
Seiberg-Witten map have been obtained in~\cite{Ghosh:2003ka,Ghosh:2004ee}.

This paper is devoted --partially-- to the study of the existence
of field configurations in noncommutative $SU(3)$ Yang-Mills theory 
that generalize the 
ordinary instanton field. We shall also analize the coupling
between massless quarks of different chirality that these
configurations give rise to and compute the corresponding  't
Hooft vertices at first order in the noncommutative parameters
$h\theta^{\mu\nu}$.  $h\theta^{\m\n}$ define the noncommutative character of 
space, for the coordinates no longer commute but satisfy 
$[X^{\m},X^{\n}]=i\,h\,\theta^{\m\nu}$. $h$ sets the noncommutative scale. Unless otherwise stated, we shall assume that
Euclidean time is commutative --i.e., that $\theta^{4\,i}=0,\; i=1,2$ and
$3$, in some reference system--; thus, upon Wick rotation the concept of 
evolution will be the ordinary one. Further, for this choice of $\theta^{\mu\nu}$, the Wick rotated action can be chosen to be at most quadratic in the first
temporal derivative of the dynamical variables at any order in the
expansion in $h\theta^{\mu\nu}$ and, thus, there is one conjugate momenta per ordinary
field. This makes it possible to use simple Lagrangian and
Hamiltonian methods to define the classical field theory and
quantize it afterwards by using elementary and standard recipes.
If time were not commutative the number of conjugate momenta will grow
with the order of the expansion in $h$ and then the Hamiltonian
formalism will have to be generalized in some way or
another~\cite{Amorim:1999mr, Gomis:2000gy}. This generalization
may affect the  quantization process in some nontrivial way and
deserves to be analyzed separately, perhaps along the lines laid
out in ref.~\cite{Amorim:1999mr}.

The layout of this article is as follows. In Section 2, we look
for --and conclude that there are none-- solutions to the $SU(N)$
noncommutative (anti)-self-duality equations that are formal power
series in $h\theta^{\mu\nu}$, with $\theta^{4i}=0$, $i=1,2$ and
$3$. Section 3 deals with the construction of field configurations
that go to the ordinary instanton as $h\theta^{\mu\nu}\rightarrow
0$ and that render stationary, at first order in
$h\theta^{\mu\nu}$, the action of noncommutative $SU(3)$
Yang-Mills theory. These field configurations will be called
first-order-in-$\theta$-deformed instantons. In Section 4, we
study the coupling between light left handed and right handed
fermions that the field configurations found in the previous
section produce and work out the appropriate 't Hooft vertices. We
do this in theories with one, two and three light
fermions. The two- and three-light fermions cases are relevant in
connection with noncommutative QCD. In the last section, we
summarize, draw conclusions and suggest how to carry on with the
program started in this paper to include corrections at second
order in $h\theta^{\m\n}$ or higher. The paper also includes five
Appendices. In Appendix A, we consider an arbitrary
$h\theta^{\m\n}$ and seek for solutions to the $SU(N)$
noncommutative (anti)-self-duality equations that come as formal
power series in $h\theta^{\mu\nu}$. The classical vacua of
non-commutative $SU(N)$ that are also formal power series in
$h\theta^{\mu\nu}$ are found in Appendix B, when time is commutative. Appendix C is devoted
to the construction at first order in $h\theta^{\mu\nu}$ of the
zero modes of the kinetic term of the quantum gauge field fluctuations
in the background of a first-order-in-$\theta$-deformed instanton.
We also compute the zero mode of the $\theta-$deformed Dirac
operator in that very background. In Appendix D, we shall show
that, when $\theta^{4i}=0$, $i=1,2,3$, no topologically nontrivial
solutions can be found as power series in
$h\theta^{\mu\nu}$ that solve the equations of motion of
noncommutative $SU(3)$ Yang-Mills theory. Several formulae used in
the paper are collected in Appendix E.

\section{Noncommutative $SU(N)$ instantons}

A noncommutative $SU(N)$ gauge field, $A_{\mu}[a_\nu]$, --see~\cite{Jurco:2001rq}-- is a self-adjoint vector field that takes values
in the enveloping algebra of the Lie algebra of $SU(N)$ and  that
is obtained from a given ordinary $SU(N)$ gauge field, $a_{\mu}$,
by means of a formal series expansion in powers of
$h\theta^{\mu\nu}$ provided by the Seiberg-Witten map. As is well
known, the Seiberg- Witten map is not unique~\cite{Asakawa:1999cu,
Brace:2001fj,Barnich:2002pb}, so that we shall call standard
Seiberg-Witten map the straightforward generalization to $SU(N)$
of the original expression introduced by Seiberg and Witten in
ref.~\cite{Seiberg:1999vs}. The standard form of the
Seiberg-Witten map reads
\begin{equation}
A_\mu=
a_\mu+\sum_{n=1}^{\infty}\;\frac{h^n}{n!}\;
\frac{d^{n-1}}{dh^{n-1}}\bigg[\frac{dA_{\mu}}{dh}\bigg]\Big\vert_{h=0}=a_\mu-\frac{h}{4}\,\theta^{\a\b}\{\aa,\Db\am+f_{\b\mu}\}\,+\,O(h^2\theta^2),
\label{cswmap}
\end{equation}
where
\begin{equation}
\frac{dA_{\mu}}{dh}=
-\frac{1}{4}\,\theta^{\a\b}\{A_{\a},\partial_{\beta}A_{\mu}+
F_{\b\mu}\}_{\star}.
\label{dotamu}
\end{equation}
$f_{\mu\nu}$ stands for the ordinary field strength. The symbol
$\star$ denotes the Moyal product of functions, i.e., $(f\star
g)(x)=f(x)\exp(\frac{i\,h}{2}\theta^{\alpha\beta}
\overleftarrow{\partial_{\alpha}}\overrightarrow{\partial_{\beta}})g(x)$,
and $\{f,g\}_{\star}= (f\star g)(x)+(g\star f)(x)$.

>From the previous $A_\mu$, one constructs the noncommutative field
strength, $F_{\mu\nu}[a_\sigma]$, as follows
\begin{equation}
\begin{array}{l}
{F_{\mu\nu}[a]\,=\,\partial_{\mu}A_\nu - \partial_{\nu}A_\mu - i[A_\mu,A_\nu]_{\star}=f_{\mu\nu}+\sum_{n=1}^{\infty}\,\frac{h^n}{n!}\;
\frac{d^{n-1}}{dh^{n-1}}\big[\frac{dF_{\mu\nu}}{dh}\big]\Big\vert_{h=0}}\\
{\phantom{F_{\mu\nu}[a]}\,=\,
f_{\mu\nu}+\frac{h}{2}\theta^{\a\b}\{f_{\mu\a},f_{\nu\b}\}-\frac{h}{4}\theta^{\a\b}\{a_{\a},(\partial_\b+\Dcal_{\b}) f_{\mu\nu}\}+O(h^2\theta^2)}.\\
\end{array}
\label{swstrengh}
\end{equation}
Here, $[A_\mu,A_\nu]_{\star}=A_\mu\star A_\nu-A_\nu\star A_\mu$ and
\begin{equation}
\frac{dF_{\mu\nu}}{dh}=\frac{1}{2}\theta^{\a\b}\{F_{\mu\a},F_{\nu\b}\}_{\star}-\frac{1}{4}\theta^{\a\b}\{A_{\a},(\partial_\b+\Dcal_{\b}) F_{\mu\nu}\}_{\star}.
\label{firstderiv}
\end{equation}

The action of a noncommutative $SU(N)$ Yang-Mills theory is given
by
\begin{equation}
    \begin{array}{l}
    S_{NCYM}=\,\frac{1}{2g^2}\,\idx \,\mathrm{Tr}\,F_{\mu\nu}\star F_{\mu\nu}=\\
    \phantom{S_{YM}}=\,\frac{1}{g^2}\,\idx\,\mathrm{Tr}\,\left[\frac{1}{2} f_{\mu\nu}f_{\mu\nu}-\frac{h}{4}\theta_{\a\b}f_{\a\b}f_{\mu\nu}f_{\mu\nu}+{h}\theta_{\a\b}\,f_{\mu\a}f_{\nu\b}f_{\mu\nu}\right]+O(h^2\theta^2).
    \end{array}
    \label{Sinst1}
\end{equation}
We shall take $a_{\mu}$ to be in the fundamental representation of
$SU(N)$. We shall only consider ordinary gauge fields, $a_{\mu}$,
such that each term in the formal expansion on the r.h.s. of
eq.~(\ref{Sinst1}) is finite. Thus, we shall impose the following
boundary condition on $a_{\mu}$:
\begin{equation}
a_{\mu}(x)\rightarrow ig(x)\partial_{\mu}g^{\dagger}(x)\,+\,O\Big(\frac{1}{\mid x \mid^2}\Big)\;\mbox{ as}\; \mid x\mid \rightarrow\infty.
\label{bound}
\end{equation}
$g(x)$ stands for an ordinary $SU(N)$ gauge transformation such that
$g(x)\rightarrow 1\;\mbox{as}\; \mid x\mid \rightarrow\infty$.

It is then postulated that  $S_{NCYM}$ governs the dynamics of our $SU(N)$
field theory on the four-dimensional noncommutative Euclidean space defined by
$[\hat{X}^\mu, \hat{X}^\nu]=i\,h \, \theta^{\mu\nu}$, with $\theta^{i4}=0$, $\forall i$.

Let us introduce the noncommutative dual field strength,
$\tilde{F}_{\mu\nu}(x)$, and its ordinary counterpart:
\begin{equation*}
\tilde{F}_{\mu\nu}\,=\,\frac{1}{2}\,\epsilon_{\mu\nu\rho\sigma}\, F_{\rho\sigma},
\quad
\tilde{f}_{\mu\nu}\,=\,\frac{1}{2}\,\epsilon_{\mu\nu\rho\sigma}\, f_{\rho\sigma}.
\end{equation*}
Then, a noncommutative $SU(N)$ field $A_{\mu}[a_\sigma]$ has
Pontrjagin index $n$ if the following equation holds
\begin{equation}
n\,=\,\frac{1}{16\pi^2}\idx\,\mathrm{Tr}\, (F_{\mu\nu}[a_\sigma]\star\tilde{F}_{\mu\nu}[a_\sigma]\,)(x).
\label{ncpontriagin}
\end{equation}
It can be shown~\cite{Martin:2005gt} that for ordinary gauge
fields satisfying the boundary conditions in eq.~(\ref{bound}),
the Pontrjagin index of $A_{\mu}[a_\sigma]$ is equal to the
Pontrjagin index of the ordinary field, $a_{\sigma}$, that defines
the former as in eq.~(\ref{cswmap}). Indeed,
\begin{equation}
\idx\,\mathrm{Tr}\, (F_{\mu\nu}[a_\sigma]\star\tilde{F}_{\mu\nu}[a_\sigma]\,)(x)
\,=\,
\idx\,\mathrm{Tr}\, f_{\mu\nu}(x) \tilde{f}_{\mu\nu}(x).
\label{swpontri}
\end{equation}

We shall say that $A_{\mu}[a_\sigma]$, defined by a given ordinary
field $a_{\mu}$ as in eq.~(\ref{cswmap}), is a noncommutative
$SU(N)$ instanton if it has Pontrjagin index --see
eq.~(\ref{ncpontriagin})-- equal to one and it is a solution to
the self-duality equation:
\begin{equation}
F_{\mu\nu}[a_{\sigma}]\,=\, \tilde{F}_{\mu\nu}[a_{\sigma}].
\label{sdequation}
\end{equation}
It is not difficult to show that every noncommutative $SU(N)$
instanton renders stationary the action in eq.~(\ref{Sinst1}).
Indeed,
\begin{equation}
 S_{NCYM}=\mp\frac{1}{2g^2}\,\idx\,\mathrm{Tr}\, F_{\mu\nu}\star\tilde{F}_{\mu\nu}+\frac{1}{4g^2}\,\idx\,\mathrm{Tr}[
(F_{\mu\nu}\pm\tilde{F}_{\mu\nu})\star (F_{\mu\nu}\pm\tilde{F}_{\mu\nu})].
\label{ncymaction}
\end{equation}

Both sides of the self-duality equation --eq.~(\ref{sdequation})--
are defined as formal power series in 
$h\theta^{\mu\nu}$ --see eq.~(\ref{swstrengh})-- around the
appropriate ordinary object: $f_{\mu\nu}$ or $\tilde{f}_{\mu\nu}$.
Hence, one would like to find solutions to this equation that are
formal series expansions in powers of $h\theta^{\mu\nu}$ around
topologically nontrivial solutions to the ordinary self-duality
equation $f_{\mu\nu}=\tilde{f}_{\mu\nu}$. We shall show below that
no such solutions exist if $\theta^{i4}=0$ in a given reference
system.

Let  $a_{\mu}$, a solution to eq.~(\ref{sdequation}), be given by
the following formal power series in $h\theta^{\mu\nu}$:
\begin{equation}
a_{\mu}[h](x)\,=\,a_{\mu}^{(0)}(x)+\sum_{k=1}^{\infty}\,h^{k}\,a_{\mu}^{(k)}(x),
\label{expasionofa}
\end{equation}
where $a_{\mu}^{(k)}(x)$ is a homogeneous polynomial in $\theta^{\mu\nu}$ of
degree $k$  whose coefficients are functions of $x$ that take values in the Lie
algebra of $SU(N)$. Then, using the expression for $F_{\mu\nu}[a_\sigma]$ on the
second line of eq.~(\ref{swstrengh}), one concludes that the following
equations hold
\begin{equation}
\begin{array}{l}
    {f^{(0)}_{\mu\nu}=\tilde{f}^{(0)}_{\mu\nu},}\\
{\Dcal^{(0)}_{\mu}a_{\nu}^{(1)}-\Dcal^{(0)}_{\nu}a_{\mu}^{(1)}+
\frac{1}{2}\theta^{\a\b}\,\{f^{(0)}_{\mu\alpha},f^{(0)}_{\nu\beta}\}
=\frac{1}{2}\epsilon_{\mu\nu\rho\sigma}\,
\Big(\Dcal^{(0)}_{\r}a_{\s}^{(1)}-\Dcal^{(0)}_{\r}a_{\s}^{(1)}+\frac{1}{2}\,\theta^{\a\b}\{
f^{(0)}_{\rho\alpha},f^{(0)}_{\sigma\beta}\}\Big),}\\
\end{array}
\label{consitency}
\end{equation}
where $\Dcal^{(0)}_{\mu}a_{\nu}^{(1)}=\partial_{\mu}a_{\nu}^{(1)}-
i[a_{\mu}^{(0)}, a_{\nu}^{(1)}]$. Since $a_{\nu}^{(1)}(x)$ takes
values in the Lie algebra of $SU(N)$, not $U(N)$, the trace over
the $SU(N)$ generators on both sides of the second equality in the
previous equation yields
\begin{equation}
\theta^{\a\b}\,f^{(0)\,a}_{\mu\alpha}f^{(0)\,a}_{\nu\beta}=\frac{1}{2}\,\theta^{\a\b}\,\epsilon_{\mu\nu\rho\sigma}f^{(0)\,a}_{\rho\alpha}f^{(0)\,a}_{\sigma\beta}.
\label{obstruction}
\end{equation}
$f^{(0)\,a}_{\mu\nu}$ stand for the components of $f^{(0)}_{\mu\nu}$ in terms
of the generators, $T^{a}$, of $SU(N)$. Now, since $\theta^{i4}=0$, we can always choose $\theta^{12}=\theta$ and $\theta^{21}=-\theta$ as only non-vanishing
components of $\theta^{\mu\nu}$. For this $\theta^{\mu\nu}$ it is not difficult
to show that the set of equations constituted by the first equality in eq.~(\ref{consitency}) and the identity in eq.~(\ref{obstruction}) is equivalent to the
following one:
\begin{displaymath}
\begin{array}{l}
{f^{(0)\,a}_{12}=f^{(0)\,a}_{34},\quad f^{(0)\,a}_{13}=-f^{(0)\,a}_{24},\quad
f^{(0)\,a}_{14}=f^{(0)\,a}_{23},}\\
\sum_a[{(f^{(0)\,a}_{12})^2+(f^{(0)\,a}_{13})^2+(f^{(0)\,a}_{14})^2]=0.}\\
\end{array}
\end{displaymath}
>From these identities one readily concludes that
$a^{(0)}_{\mu}$ has vanishing field strength, so that it is gauge
equivalent to the vanishing  gauge field.

We have shown so far that when the standard Seiberg-Witten map
--defined in eqs.~(\ref{cswmap}) and~(\ref{dotamu})-- is employed
to define the noncommutative field strength $F_{\mu\nu}$, the
self-duality equation --eq.~(\ref{sdequation})-- has no solution
of the type displayed in eq.~(\ref{expasionofa}). We shall show
next that this state of affairs remains unaltered for the most
general type of Seiberg-Witten map. At first order in
$h\theta^{\mu\nu}$, the most general expression for the
Seiberg-Witten map reads
\begin{equation}
A_{\mu}=a_\mu-\frac{h}{4}\theta^{\a\b}\{\aa,\Db\am+f_{\b\mu}\}+
\kappa_{1}\,h\theta^{\a\b}\,\Dcal_{\mu} f_{\a\b}+
\kappa_{2}\,h\theta^{\a\b}\,\Dcal_{\mu} [a_{\a},a_{\b}]+
\kappa_{3}\,h\theta_{\mu}^{\phantom{\mu}\b}\Dcal^{\nu} f_{\nu\b}+ O(h^2\theta^2),
\label{SWgeneral}
\end{equation}
where $\kappa_{i},\;i=1,2,3$ are arbitrary real numbers. The noncommutative
field strength for the previous noncommutative gauge field is given by
\begin{displaymath}
\begin{array}{l}
{F_{\mu\nu}=
f_{\mu\nu}+\frac{h}{2}\theta^{\a\b}\{f_{\mu\a},f_{\nu\b}\}-\frac{h}{4}\theta^{\a\b}\{a_{\a},(\partial_\b+\Dcal_{\b}) f_{\mu\nu}\}
-i\kappa_{1}\,h\theta^{\a\b}[f_{\m\n},f_{\a\b}]}\\
{\phantom{F_{\mu\nu}=}-i\kappa_{2}\,h\theta^{\a\b}[f_{\m\n},[a_{\a},a_{\b}]]
-\kappa_{3}\,h\big(\theta_{\mu}^{\phantom{\mu}\b}\Dcal_{\nu}\Dcal^{\r} f_{\r\b}
-\theta_{\nu}^{\phantom{\nu}\b}\Dcal_{\mu}\Dcal^{\r} f_{\r\b}\big)+O(h^2\theta^2).}
\end{array}
\end{displaymath}
Substituting this expression in both sides of
eq.~(\ref{sdequation}), one concludes that eq.~(\ref{obstruction})
is not modified by the new terms in the previous $F_{\mu\nu}$.
Hence, no solutions to the noncommutative self-duality equation
can be found by using formal powers series in $h\theta^{\mu\nu}$
around ordinary fields with non-vanishing instanton number.

It is clear that the result we have obtained for the self-duality
equation also carries over to the anti-self-duality equation:
\begin{displaymath}
F_{\mu\nu}=-\tilde{F}_{\mu\nu}.
\end{displaymath}

The reader may wonder what would the situation be had we assumed
an arbitrary $\theta^{\mu\nu}$. In Appendix A  we show that the
self-duality equation defined by the standard Seiberg-Witten map
has  topologically nontrivial solutions  of the type in
eq.~(\ref{expasionofa}) if, and only if, $\theta^{\mu\nu}$ is
self-dual:
$\theta_{\mu\nu}=\frac{1}{2}\epsilon_{\m\n\r\s}\theta^{\r\s}$.
Actually, these solutions are the ordinary  gauge field
configurations that are self-dual. Analogously, the
anti-self-duality equation has solutions of the type in
eq.~(\ref{expasionofa}) with non-vanishing Pontrjagin number if,
and only if, $\theta^{\mu\nu}$ is anti-self-dual:
$\theta_{\mu\nu}=-\frac{1}{2}\epsilon_{\m\n\r\s}\theta^{\r\s}$.
These solutions are ordinary field configurations that are
anti-self-dual. In other words, if $\theta^{\mu\nu}$ is self-dual,
the standard Seiberg-Witten map in eq.~(\ref{cswmap}) maps
ordinary (multi-)instantons into noncommutative (multi-)instantons
and if $\theta^{\mu\nu}$ is anti-selfdual, the standard
Seiberg-Witten map maps ordinary (multi-)anti-instantons into
noncommutative (multi-)anti-instantons.

\section{First-order-in-$\theta$-deformed instantons}

In this section we shall look for ordinary field configurations,
$a_{\m}[h](x)$, that render stationary the action of Euclidean
noncommutative $SU(3)$ Yang-Mills theory up to first order in $h\theta^{\m\n}$
and that admit a formal series expansion in powers of
$h\theta^{\m\n}$. We shall further assume that these field
configurations have a smooth dependence on the space coordinates,
that they satisfy the boundary conditions in eq.~(\ref{bound}) and
that they go to the ordinary instanton as $h\rightarrow 0$.

 From eq.~(\ref{Sinst1}), one derives the following equation of motion up to
first order in $h\theta^{\m\n}$:
\begin{equation}
    \Tr\, T^a\left[\Dcal_\mu f_{\mu\nu}+h\theta^{\a\b}\left\{f_{\mu\b},-\Dcal_\a f_{\nu\mu}+\frac{1}{2}\Dcal_\nu f_{\a\mu}\right\}\right]=0+O(h^2\theta^2).
\label{motionuptoh1}
\end{equation}

Now, any ordinary $SU(3)$ instanton, $a^{\rm{oinst}}_\mu$, has the
form
\begin{equation*}
a^{\rm{oinst}}_\mu(x)\,=\,U a^{\bp}_\mu(x) U^{\dagger},
\end{equation*}
where $U$ is an arbitrary rigid $SU(3)$ transformation and
$a^{\bp}_\mu(x)$ denotes the upper-left-hand  corner embedding in $SU(3)$ 
of the ordinary $SU(2)$ instanton, which can be written as
\begin{equation}
 a^{\bp}_\mu(x)\,=\,\eta_{a\mu\nu}\frac{(x-x_0)_\nu}{(x-x_0)^2+\rho^2}\,T^a
\label{ordbpst}
\end{equation}
in the regular gauge. $T^{a}$, $a=1,2,3$ denote the
upper-left-hand corner embedding of the $SU(2)$ generators into the
$SU(3)$ generators and $\eta_{a\mu\nu}$ stands for the self-dual
't Hooft symbols~\cite{'tHooft:1976fv}.

Since eq.~(\ref{motionuptoh1}) is invariant under $SU(3)$ transformations,
we shall find first a solution to it of the form
\begin{equation}
a_\mu[h]=a^{\bp}_\mu+h b_\mu+O(h^2\theta^2).
\label{ansatz}
\end{equation}
Then, we shall apply to this solution an arbitrary rigid $SU(3)$
transformation. This type of solutions will be called
first-order-in-${\theta}$-deformed instantons. $b^{\mu}$ in
eq.~(\ref{ansatz}) is an $SU(3)$ Lie-algebra-valued smooth vector
field which is linear in $\theta^{\m\n}$ and vanishes rapidly enough
at infinity.

Let us substitute the previous $a_\mu[h]$ in
eq.~(\ref{motionuptoh1}) and discard any contribution of order
$h^2\theta^2$. Of course, the order $h^0$ contribution thus obtained is
satisfied by construction: $\Dcal^{\bp}_\mu f^{\bp}_{\mu\nu}=0$.
The order $h\theta^{\m\n}$ contribution yields a non-homogeneous equation for
$b_\mu$:
\begin{equation}
\begin{array}{l}
    -i \Tr\,T^a [b_\mu,f^{\bp}_{\mu\nu}]+\Tr\, T^a[\Dcal^{\bp}_\mu\Dcal^{\bp}_\mu b_\nu-\Dcal^{\bp}_\mu\Dcal^{\bp}_\nu b_\mu]=\\
    \frac{1}{2}\,\Tr\,T^a \theta^{\a\b}\{f^\bp_{\mu\b},\Dcal^\bp_\a f^\bp_{\nu\mu}+\Dcal^\bp_\mu f^\bp_{\nu\a}\}.
\end{array}
\label{ecmov}
\end{equation}\par
    The action of $[T^a, ],\,a\in\,\{1,2,3\}$ over the generators of $SU(3)$ defines
four irreducible representations of $SU(2)$: a spin one
representation acting on the linear span of $\{T^1,T^2,T^3\}$, two spin $1/2$
representations acting on  the linear span of $\{T^4,T^5,T^6,T^7\}$ and a singlet
acting on $T^8$. On the other hand, if $a,b\in\{1,2,3\}$, we have
$\{T^a,T^b\}=\frac{1}{3}\delta^{ab}\id+d_{abc}T^c|_{a,b\in\{1,2,3\}}=\frac{1}{3}\delta^{ab}\id+\frac{1}{\sqrt{3}}\delta^{ab}T^8$.
This last identity implies that the r.h.s. of eq.~(\ref{ecmov}) is
non-zero only for $a=8$. Let us  express $b_\nu$ as
\begin{equation}
    b_\nu=b^{(1\dots7)}_\nu+b^8_{\n}T^8,\quad b^{(1\dots7)}_\nu=\sum_{c=1}^7
b^{c}_\nu T^c.
\label{bnu}
\end{equation}
Then, taking into account that the action of $[T^a, ],\,a\in\,\{1,2,3\}$,
does not mix the irreducible representations mentioned above, one
concludes that the equation of motion for $b^{(1\dots7)}_\nu$
decouples from that of $b^8_\nu$. For $b^{(1\dots7)}_\nu$ the
equation of motion reads
\begin{equation}
    (\Dcal^\bp_\mu(\Dcal^\bp_\mu b^{(1\dots7)}_\nu-\Dcal^\bp_\nu b^{(1\dots7)}_\mu)-i[b^{(1\dots7)}_\mu,f^\bp_{\mu\nu}]) ^a=0,\,\,\,\,a\in{1,\dots,7},
\label{b17motion}
\end{equation}
whereas for $b^8_\nu$, we have the following non-homogeneous
equation:
\begin{equation}
    (\square\,\delta_{\m\n}-\Dm\Dn)\, b^8_\nu=\sum_{a}\frac{\theta^{\a\b}}{2\sqrt{3}}[f^{\bp\,a}_{\mu\b}(\Dcal^{\bp}_\a f^{\bp}_{\nu\mu}+\Dcal_\mu^{\bp}f^\bp_{\nu\a})^a].
    \label{b8}
\end{equation}

Let us first solve this second equation. The most general
solution to eq.~(\ref{b8}) is of the form
$b^8_\mu=b^{8\,(hom)}_\mu+b^{8\,(part)}_\mu$, $b^{8\,(part)}_\mu$
being a particular solution to it and $b^{8\,(hom)}_\mu$ denoting
the most general solution to the corresponding homogeneous
equation
\begin{equation*}
    (\square\, \delta_{\m\n}-\Dm\Dn)\, b^{8\,(hom)}_\nu=0.
\end{equation*}
Any solution to the latter equation which is smooth and vanishes at
infinity reads $b^{8\,(hom)}_\mu =
\partial_{\mu}\phi$, where $\phi$ is an appropriate function.
Recalling that $T^8$ commutes with $T^a,\,a\in\{1,2,3\}$, one
concludes that this $b^{8\,(hom)}_\mu$ can always be generated
by applying to $a_\mu[h]$ in eq.~(\ref{ansatz}) the following
gauge transformation: $g(x)=e^{ih\phi(x)T^8}$. We are thus left
with the problem of finding a particular solution,
$b^{8\,(part)}_\mu$, to eq.~(\ref{b8}).

Let us assume that $b^{8\,(part)}_\mu$ satisfies the
transversality condition $\partial_\mu b^{8\,(part)}_\mu=0$ and
has the following form:
\begin{equation}
 b^{8\,(part)}_\mu   =\theta_{\mu\nu}(x-x_0)_\nu f[(x-x_0)^2]-\,\tilde{\theta}_{\mu\nu}(x-x_0)_\nu g[(x-x_0)^2],\quad\tilde{\theta}_{\mu\nu}=\frac{1}{2}\epsilon_{\mu\nu\rho \sigma}\theta_{\rho\sigma},
\label{aparticular}
\end{equation}
where $x_0$ is the centre of the ordinary BPST  instanton. Then,
eq.~(\ref{b8}) boils down to  the following equation to be
satisfied by  $f[(x-x_0)^2]$ and $g[(x-x_0)^2]$:
\begin{equation*}
3y'+(x-x_0)^2y''=-\frac{48\rho^4}{\sqrt{3}[(x-x_0)^2+\rho^2]^5},\quad\quad
y=f,\,g.
\end{equation*}
A solution to this equation that is smooth and vanishes at
infinity is given by
\begin{equation*}
y(x)=f[(x-x_0)^2]=g[(x-x_0)^2]=\frac{2}{\sqrt{3}}\frac{r^2+3\rho^2}{(r^2+\rho^2)^3},
\end{equation*}
where $r^2=(x-x_0)^2$. Substituting the previous result in
eq.~(\ref{aparticular}), one finally gets  that, modulo gauge
transformations, the solution, $b^8_{\mu}$, to eq.~(\ref{b8}) that
is smooth and vanishes at infinity reads
\begin{equation}
 b^8_{\mu}=b^{8\,(part)}_\mu   =\frac{2}{\sqrt{3}}\,(\theta_{\mu\nu}(x-x_0)_\nu
 -\tilde{\theta}_{\mu\nu}(x-x_0)_\nu)\,\frac{r^2+3\rho^2}{(r^2+\rho^2)^3}.
\label{finalb8}
\end{equation}
Again, $r^2=(x-x_0)^2$.

Let us now solve eq.~(\ref{b17motion}). Using the fact that $f_{\mu\nu}^\bp=\tilde{f}^\bp_{\mu\nu}$ and the relation
$[f_{\mu\nu},]=i[\Dcal_\mu,\Dcal_\nu]$, this equation can be turned into the
following equality:
\begin{equation}
\big(\Dcal^\bp_\mu\big\{[\Dcal^\bp_\mu b^{(1\dots7)}_\nu-\Dcal^\bp_\nu b^{(1\dots7)}_\mu]-\frac{1}{2}\epsilon_{\m\n\r\s}
[\Dcal^\bp_\r b^{(1\dots7)}_\s-\Dcal^\bp_\s b^{(1\dots7)}_\r]\big\}\big)^a=0,\,\,a\in{1,\dots,7}.
\label{b17}
\end{equation}
We shall  show next that any smooth $b^{(1\dots7)}_\mu$ that vanishes
at infinity is a solution to the previous equation if, and only if, it
solves the following equality:
\begin{equation}
[\Dcal^\bp_\mu b^{(1\dots7)}_\nu-\Dcal^\bp_\nu b^{(1\dots7)}_\mu]-\frac{1}{2}\epsilon_{\m\n\r\s}
[\Dcal^\bp_\r b^{(1\dots7)}_\s-\Dcal^\bp_\s b^{(1\dots7)}_\r]=0.
\label{zeromodeseq}
\end{equation}
Let $\Omega_{\m\n}$ denote the left hand side of the previous equation. Then
eq.~(\ref{b17}) reads
\begin{equation}
 \Dcal^\bp_\mu\Omega_{\m\n}=0.
\label{eqomega}
\end{equation}
Since $\Omega_{\m\n}$ is an anti-symmetric and anti-self-dual
object, it has the following representation in terms of a 
spinorial object $\Omega^{\a\b}$:
$\Omega_{\m\n}=i(\s_{\m\nu})_{\a\b}\,\Omega^{\a\b}$, where
$\s_{\m\nu}=-\frac{1}{4i}(\s_{\m}\bar{\s}_\n-\s_{\n}\bar{\s}_\m)$
with $\s_{\m}=(\vec{\s},i)$ and $\bar{\s}_{\m}=(-\vec{\s},i)$. In
terms of $\Omega^{\a\b}$,  eq.~(\ref{eqomega}) reads
\begin{equation*}
\tr\big(\s_\n \Dcal^\bp_\mu
\bar{\s}_\m \Omega^{\top}\big)=0,
\end{equation*}
 where ``$\tr$'' stands for the trace over the spinor indices and $\Omega^{\top}$
is the transpose of $\Omega$ with regard to the latter indices.
Since $\{\s_\n\}$ is an orthogonal basis of the $2\times 2$ matrices,
the previous equation is equivalent to
\begin{equation*}
\Dcal^\bp_\mu
\bar{\s}_\m \Omega^{\top}=0.
\end{equation*}
Applying $\Dcal^\bp_\nu
\s_\n$ to this equation, one concludes that
\begin{equation*}
(\Dcal^\bp)^2\,\Omega^{\top} =0.
\end{equation*}
Indeed, just take into account that $\s_\m
\bar{\s}_\n=-\delta_{\m\n}-2i\s_{\m\n}$ and that
$\s_{\m\n}f^{\bp}_{\m\n}=0$. The latter equality is a consequence
of $\s_{\m\n}$ being anti-self-dual and $f^{\bp}_{\m\n}$ being
self-dual. Now, $(\Dcal^\bp)^2$ is a positive definite operator,
so that it has no normalizable non-vanishing eigenvectors with zero
eigenvalue --see~\cite{Coleman:1978ae}. Hence,
\begin{equation*}
\Omega^{\a\b}=0.
\end{equation*}
Recalling that $\Omega_{\m\n}=i(\s_{\m\nu})_{\a\b}\,\Omega^{\a\b}$,
we conclude that $\Omega_{\m\n}(x)$ --smooth anti-symmetric
anti-self-dual object that vanishes at infinity-- satisfies
eq.~(\ref{eqomega}) if, and only if, $\Omega_{\m\n}(x)=0$. We have
thus shown that solving eq.~(\ref{b17}) is equivalent to solving
eq.~(\ref{zeromodeseq}) for smooth functions that vanish at
infinity rapidly enough. Now, eq.~(\ref{zeromodeseq}) is the
equation for the zero modes of the ordinary $SU(3)$
instanton~\cite{Bernard:1977nr}. Hence, the $b^{(1\dots7)}_\nu$ we
are looking for are linear combinations of those zero modes with
coefficients that depend linearly on $\theta^{\m\nu}$, and can thus
be obtained by deforming infinitesimally the collective
coordinates of a given $SU(3)$ ordinary instanton. Since this
deformation yields another $SU(3)$ instanton, we shall set,
without loss of generality, $b^{(1\dots7)}_\nu$ to zero.
Substituting both this result and eq.~(\ref{finalb8}) in
eq.~(\ref{bnu}), and the result so obtained, in turn, in eq.~(\ref{ansatz}),
one gets the most general first-order-in-$\theta$-deformed
instanton, $a_{\m}[h]^{({\rm gen})}$, in the regular gauge:
\begin{equation}
a_{\m}[h]^{({\rm gen})}(x)\,=\,Ua_{\m}[h](x)U^{\dagger}.
\label{genertalforin}
\end{equation}
Here, $U$ is an arbitrary rigid $SU(3)$ transformation that does
not leave $a_{\m}[h]$ invariant and $a_{\m}[h]$ is given by
\begin{equation}
 a_{\m}[h]=a^{\bp}_\mu\,+\,h\,T^8\frac{2}{\sqrt{3}}\,[\theta_{\mu\nu}
 -\tilde{\theta}_{\mu\nu}](x-x_0)_\nu\,\frac{r^2+3\rho^2}{(r^2+\rho^2)^3},
\label{fordertheta}
\end{equation}
where $a^{\bp}_\mu$ is defined in eq.~(\ref{ordbpst}) and
$r^2=(x-x_0)^2$.

It can be shown that  $a_{\m}[h]$ has instanton number equal to 1 and that its
contribution to the noncommutative Yang-Mills action in eq.~(\ref{ncymaction})
reads
\begin{equation*}
    S_{NCYM}=\frac{8\pi^2}{g^2}\,+\,O(h^2\theta^2).
\end{equation*}
Hence, at the order we are working --first order in $h\theta^{\m\n}$-- $a_{\m}[h]$ gives no
correction to the famous value --$\frac{8\pi^2}{g^2}$-- of the corresponding ordinary theory.

>From eq.~(\ref{genertalforin}), one learns that, at first order
in $h\theta^{\m\n}$,  the moduli space of non-commutative $SU(3)$
Yang-Mills theory has dimension 12 for the $k=1$ instanton sector.
Indeed, as in the ordinary case, there are 12 collective
coordinates that parametrise $a_{\m}[h]^{(\rm gen)}(x)$ in
eq.~(\ref{genertalforin}): $\rho$, $x_0^{\m}$ and the seven angles
of the coset space $SU(3)/U(1)$.

In the next section and in Appendix C, we shall use  our generic
first-order-in-$\theta$-deformed instanton  in the singular gauge, which we shall denote by $a_{\m}^{({\rm gsing})}$. $a_{\m}^{({\rm
gsing})}$ is given by
\begin{equation}
\begin{array}{l}
{a_{\m}^{({\rm gsing})}(x)=Ua_{\m}(x)^{({\rm
sing})}U^{\dagger},}\\
{a_{\m}^{({\rm sing})}(x)=
\overline{\eta}_{a\mu\nu}\,\frac{\r^2\,(x-x_0)_\nu}{(x-x_0)^2[(x-x_0)^2+\rho^2]}\,\tau^a+\frac{2h}{\sqrt{3}}\,
(\theta-\tilde{\theta})_{\mu\a}(x-x_0)_\a\frac{(x-x_0)^2+3\rho^2}{((x-x_0)^2+\rho^2)^3}\,T^8.}\\
\end{array}
\label{singinst}
\end{equation}
$\tau^a$, $a=1,2$ and $3$, stand for the upper-left-hand corner
embedding of the $SU(2)$ generators in the generators of $SU(3)$,
both sets of generators being in their fundamental
representations. $a_{\m}^{({\rm sing})}(x)$ is obtained by
applying to $a_{\m}[h](x)$ in eq.~(\ref{fordertheta}) the
following $SU(3)$ gauge transformation $g(x)=\frac{i\tau_\mu^+
(x-x_0)_\mu}{\sqrt{(x-x_0)^2}}$, where
$\tau^+_\mu=(\overrightarrow{\tau},-i)$ and
$\overrightarrow{\tau}=(\tau^1,\tau^2,\tau^3)$.

 We shall close this section by making the connection between the
first-order-in-$\theta$-deformed instanton in
eq.~(\ref{fordertheta}) and the classical vacua of noncommutative
$SU(3)$ Yang-Mills theory. In ordinary $SU(3)$ Yang-Mills theory
the instanton interpolates --along Euclidean time--  between a classical 
vacuum in the
distant past that has winding number $n$ and a classical vacuum
in the distant future with winding number equal to $n+1$. We shall
see below that the same type of phenomenon occurs when our
first-order-in-$\theta$-deformed instanton is at work. Since the
phenomenon in question  is most easily exhibited  in the Euclidean temporal
gauge, we shall perform a gauge transformation so that our
first-order-in-$\theta$-deformed instanton satisfies the gauge
condition $a_4[h]=0$. Our first-order-in-$\theta$-deformed
instanton in the temporal gauge, $a_4[h]=0$, reads thus
\begin{equation*}
 a_{i}[h]^{temporal}(\vec{x},\tau)=g(\vec{x},\tau)a_{i}[h](x)g(\vec{x},\tau)^{\dagger}+
ig(\vec{x},\tau)\partial_{i} g(\vec{x},\tau)^{\dagger},\quad i=1,2,3,
\end{equation*}
where $x=(\vec{x},\tau)$, $a_{i}[h](x)$ is given by the r.h.s of eq.~(\ref{fordertheta}) and
\begin{equation*}
g(\vec{x},\tau)\,=\,\big(e^{i\int_{-\infty}^{\tau}\,a_{4}[h](\vec{x},t)\,dt}\big)g_{-}(\vec{x}).
\end{equation*}
$a_{4}[h](\vec{x},t)$ is defined by the r.h.s. of
eq.~(\ref{fordertheta}) and $g_{-}(\vec{x})\epsilon\, SU(3)$ is
such that $g_{-}(\mid \vec{x}\mid\rightarrow\infty)=1$. We see
that $a_{i}[h]^{temporal}(\tau=-\infty,\vec{x})\equiv a_{i}^{-}(\vec{x})=
ig_{-}(\vec{x})\partial_{i}g_{-}(\vec{x})^{\dagger}$ and that
$a_{i}[h]^{temporal}(\tau=+\infty,\vec{x})\equiv a_{i}^{+}(\vec{x})=
ig_{+}(\vec{x})\partial_{i}g_{+}(\vec{x})^{\dagger}$, with
\begin{equation*}
g_{+}(\vec{x})=\exp\Big[\frac{-i\pi\vec{x}\vec{\sigma}}{\sqrt{\vec{x}^2+\rho^2}}\Big]\;
\exp[-i\pi\phi^{8}(\vec{x})T^8]\;g_{-}(\vec{x}),\quad\quad
\phi^{8}(\vec{x})=h\frac{2}{\sqrt {3}}\,\tilde{\theta}_{0i}x_{i}\,
\frac{2\vec{x}^2+5\r^2}{(\vec{x}^2+\r^2)^{5/2}}.
\end{equation*}
 Now, it can be shown --see Appendix B-- that, for commutative
time and in the temporal gauge, all classical vacua,    $a_{i}[h](\vec{x})$,
of the noncommutative Yang-Mills theory
that admit a formal series expansion in powers of $h\theta^{\m\nu}$ are of
the form $a_{i}[h](\vec{x})=ig(\vec{x})\partial_{i}g(\vec{x})^{\dagger}$
--$g(\vec{x})\epsilon\, SU(3)$. Now, if
$a_{i}^{-}(\vec{x})$ has winding number equal to $n$, $a_{i}^{+}(\vec{x})$ has winding number equal to $n+1$: notice that the famous hedgehog matrix occurs in the definition of
 $g_{+}(\vec{x})$ and that $\exp(-i\pi\phi^{8}(\vec{x})T^8)$ has vanishing winding number. We have thus shown that for noncommutative $SU(N)$ Yang-Mills theory, when time is commutative, our first-order-in-$\theta$-deformed instanton  field connects along Euclidean time a classical
vacuum in the distant past  with a classical vacuum in the distant
future, the latter having a winding number  which is one unit
greater than the former's. This transition cannot be accomplished
by continuous evolution along the classical trajectories --i.e., solutions of
the equations of motion on noncommutative Minkowski space-time--
since it involves a change of the winding number. The phenomenon,
as in ordinary Minkowski space-time, is a genuine quantum effect:
the transition is realised by tunnelling between the two vacua. We
shall analyze this tunnel effect in the next section.

\section{Vacuum to vacuum transition and `t Hooft vertices}

For Euclidean signature and at first order in $h\theta^{\m\n}$,
the action of non-commutative $SU(3)$ gauge theory with $n_f$
Dirac fermions is obtained by adding to $S_{NCYM}$ in
eq.~(\ref{Sinst1}) the fermionic action $S_{F}$, which is given by:
\begin{equation}
S_F=-\sum_{f=1}^{n_f}\idx \psib_{f}\left[{\cal
K}[a_\m]+im_f\right]\psi_f.
\label{fermact}
\end{equation}
Here, ${\cal K}$ denotes the following $\theta$-deformation of the
ordinary Dirac operator $i\Dirac[a_\m]$:
\begin{equation}
{\cal
K}[a_\m]=i\Dirac[a_\m]-\frac{ih}{2}\theta_{\alpha\beta}\g_\rho
f_{\rho\alpha}
D_\beta[a_\m]+\frac{ih}{8}\theta_{\a\b}\g_\mu(\Dcal_\mu
f_{\a\b})[a_\r].
 \label{defop}
\end{equation}
This operator has --at least in  perturbation theory in
$h\theta^{\m\n}$-- a discrete spectrum for gauge field
configurations such as $a_\m^{({\rm gsing})}$ in
eq.~(\ref{singinst}). See ref.~\cite{Martin:2005jy} for further
details.

Let us denote by $\tilde{a}_{\m}^a$ the quantum fluctuations
around the first-order-in-$\theta$-deformed instanton in the
singular gauge, $a_\m^{({\rm gsing})}$: $a^a_{\m}=a_{\m}^{a\,({\rm
gsing})}+\tilde{a}_{\m}^a$. Then, in the
first-order-in-$\theta$-deformed instanton transition, the vacuum
to vacuum amplitude for the noncommutative gauge theory with
action $S=S_{NCYM}+S_{F}$ is given, at one-loop level, by the
following path integral in the background-field gauge:
\begin{equation}
\begin{array}{l}
{\langle vac
,n=1|vac,n=0\rangle=e^{-\frac{8\pi^2}{g^2}}\,\int\,d\gamma\,J(\gamma)\,\int\,d\tilde{a}_{\m}^a\,
\int\,d\bar{c}^a\,dc^a 
\int\,\prod_f\;d\psib_f\,d\psi_{f}}\\
{\phantom{\langle vac} e^{-\frac{1}{2}\idx\,\tilde{a}_{\m}^a
{\cal M}^{ab}_{\m\n}[a_{\r}^{({\rm
gsing})}]\,\tilde{a}_{\n}^b+\idx\,\bar{c}^a{\cal
M}^{ab}_{gh}[a_{\m}^{({\rm gsing})}]\,c^b+\sum_{f=1}^{n_f}\idx
\psib_{f}\left[{\cal K}[a_{\m}^{({\rm
gsing})}]+im_f\right]\psi_f}}\\
{\phantom{<0}=e^{-\frac{8\pi^2}{g^2}}\,\Big(\det'\big({\cal
M}^{ab}_{\m\n}[a_{\r}^{({\rm
gsing})}]\big)\Big)^{-1/2}\,\det\big({-\cal
M}^{ab}_{gh}[a_{\m}^{({\rm gsing})}]\big)\,\prod_{f=1}^{n_f}\,
\det\big(-{\cal K}[a_{\m}^{({\rm gsing})}]-im_f\big).}\\
\end{array}
\label{pathtrans}
\end{equation}
Let us spell out now what the new symbols in the previous
identity stand for. $|vac,n=0\rangle$ and $|vac,n=1\rangle$
denote  vacua corresponding, respectively, to  gauge field
configurations with winding number $n=0$ and $n=1$, these vacua being
connected by our first-order-in-$\theta$-deformed instanton.
$\gamma$ denotes the collective coordinates of $a_{\m}^{({\rm
gsing})}$, namely: its size $\rho$, its  center $x_0$ and its
orientation --given by $U$, a rigid
$SU(3)$ transformation-- in the Lie algebra of $SU(3)$.
$J(\gamma)$ is the collective coordinates
Jacobian, which is computed, in Appendix C, from the zero modes of
the operator ${\cal M}^{ab}_{\m\n}[a_{\r}^{({\rm gsing})}]$
defined below. The fields $c^a$ and $\bar{c}^a$ are the ghost
fields introduced in the gauge-fixing procedure. The operators
${\cal M}^{ab}_{\m\n}[a_{\r}^{({\rm gsing})}]$ and ${\cal
M}^{ab}_{gh}[a_{\m}^{({\rm gsing})}]$ are defined by the following
identities:
\begin{equation}
\begin{array}{l}
{{\cal M}^{ab}_{\m\n}[a_{\s}^{({\rm gsing})}]=
\frac{\delta^2S_{NCYM}}{\delta{a^a_\mu}\delta{a^b_\nu}}\vert_{a^a_{\s}=a_{\s}^{a\,({\rm
gsing})}}+\Dcal^{ac}_{\m}[a_{\s}^{({\rm gsing})}]\Dcal^{cb}_{\n}[a_{\s}^{({\rm gsing})}],}\\[10pt]
{{\cal M}^{ab}_{gh}[a_{\s}^{({\rm gsing})}]=-\big(\Dcal^2[a_{\s}^{({\rm gsing})}]\big)^{ab}.}\\
\end{array}
\label{quadoper}
\end{equation}
Here,  $\Dcal^{ab}_{\m}[a_{\r}^{({\rm
gsing})}]=\partial_{\m}\delta^{ab}-f^{abc}a _{\m}^{c\,({\rm
gsing})}(x)$ and  $S_{NCYM}$ is given in eq.~(\ref{Sinst1}). Let
us finally note that $\det'{\cal M}^{ab}_{\m\n}[a_{\r}^{({\rm
gsing})}]$ indicates that the zero modes of ${\cal
M}^{ab}_{\m\n}[a_{\r}^{({\rm gsing})}]$ --see Appendix C-- are to
be left out when computing the determinant.

Now, as shown in Appendix C, ${\cal K}[a_{\m}^{({\rm gsing})}]$
has a zero mode, at least at first order in $h\theta^{\m\n}$. Hence,
\begin{equation}
\det\big({-\cal K}[a_{\m}^{({\rm
gsing})}]-im_f\big)=-im_f\,\prod_{\lambda > 0}\,(-\lambda^2-m_f^2),
\label{eigenprod}
\end{equation}
where $\lambda$ denotes a generic positive eigenvalue of ${\cal
K}[a_{\m}^{({\rm gsing})}]$. To obtain eq.~(\ref{eigenprod}), we
have taken into account that the non-zero eigenvalues of ${\cal
K}[a_{\m}^{({\rm gsing})}]$ come in pairs $(\lambda, -\lambda)$.
The spectrum of ${\cal K}[a_{\m}^{({\rm gsing})}]$ is discrete, at
least in the perturbative expansion in $h\theta^{\m\n}$, due to
the fast fall-off of $a_{\m}^{({\rm gsing})}$ at infinity.

That the vacuum to vacuum amplitude in eq.~(\ref{pathtrans})
vanishes, or nearly vanishes, when massless, or nearly massless,
quarks couple to the first-order-in-$\theta$-deformed instanton
can be seen as a consequence of the $U(1)_{A}$ anomaly. Indeed,
using the results in refs.~\cite{Martin:2005gt, Martin:2005jy},
one concludes that in the first-order-in-$\theta$-deformed
instanton transition the chiral charge associated to each massless
flavour, $f$, changes compulsorily by two units: the
first-order-in-$\theta$-deformed instanton turns a left handed
quark into a right handed quark. This selection rule would be
broken by a non-zero $\langle vac ,n=1|vac,n=0\rangle$.  Since the
first-order-in-$\theta$-deformed instanton turns a left-handed
massless quark into a right-handed quark with the same
flavour, to obtain non-zero amplitudes one must insert enough
pairs of quark-anti-quark fields between $|vac,n=0\rangle$ and
$\langle vac ,n=1|$. Indeed, the quark propagator of the flavour
$f$ in the first-order-in-$\theta$-deformed instanton reads
\begin{equation}
\langle \psi_f (x) \psib(y)_f \rangle^{(\theta definst)} =-
\frac{\psi_0(x)\psi_0(y)^{\dagger}}{im_f}\,-\,\sum_{\lambda \neq
0}\,\frac{\psi_\lambda (x)\psi^{\dagger}_\lambda (y)}{\lambda +
im_f}, \label{propinst}
\end{equation}
where $\psi_0(x)$ stands for the zero-mode of ${\cal
K}[a_{\m}^{({\rm gsing})}]$ worked out in Appendix C and
$\psi_\lambda (x)$ denotes generically the remaining
eigenfunctions of this operator. Suppose now that the masses of
the $n_{f}$ quark flavours are taken to zero.  Then, in this
chiral limit, the Green function $\langle\prod_{f=1}^{n_{f}}\psi_f
(x) \psib(y)_f \rangle^{(\theta definst)}$ has a non-vanishing
value, for the pole at $m_f=0$ of the propagator in
eq.~(\ref{propinst}) cancels the contribution linear in $m_f$ appearing in
the determinant in eq.~(\ref{eigenprod}).

As in the ordinary case~\cite{'tHooft:1976fv}, the coupling
between left-handed and right-handed massless quarks through the
first-order-in-$\theta$-deformed instanton can be mimicked by
using an effective Lagrangian. This coupling  does not occur at
any order in the perturbative expansion in powers of the coupling
constant and the effective Lagrangian, ${\cal
L}_{eff}=\sum_{n=0}^{n_{f}}\,{L}_{2n}$, that simulates it is
a sum of non-local interactions --called 't Hooft vertices--, each
involving $2n$ fermions. In these non-local interactions quarks
are emitted or absorbed in the zero-mode wave function $\psi_0
(x)$. The contribution ${L}_{2n}$ matches, as $m_f\rightarrow
0$ --$f=1,...,n_{f}$--, the leading contribution to the amputated Green
function obtained from $\langle \prod_{f'=1}^{n}\,\psi_{f'}
(x_{f'}) \psib(y_{f'})_{f'} \rangle^{(\theta definst)}$. The amputation is
to be carried out with the Dirac free propagator, and
$\{\psi_{f'} (x)\}$ stands for any set of $n$ --with $n\leq n_f$--
fermion fields.
Now, it is further assumed that the previous Green function is
normalized to the vacuum to vacuum amplitude in the perturbation
theory background $a_{\mu}=0$. Below we shall work out this
effective Lagrangian for one, two and three nearly massless
flavours --i.e., $n_{f}=1,2$ and $3$-- to obtain the first order in
$\theta$ corrections to the ordinary results obtained in
refs.~\cite{Shifman:1979uw,Nowak:1988bh, Diakonov:2002fq}.

\subsection{One light flavour}

In this case we need to compute the leading contribution as, say,
$m_1=m\rightarrow 0$, to $\langle vac ,n=1|vac,n=0\rangle/\langle
vac ,n=0|vac,n=0\rangle$ and $\langle
\psi(x)\psi^{\dagger}(y)\rangle^{(\theta definst)} /\langle vac
,n=0|vac,n=0\rangle$. $\psi(x)$ denotes the  field of the light 
quark. It turns out that the contribution  to $\langle vac
,n=1|vac,n=0\rangle/\langle vac ,n=0|vac,n=0\rangle$ which is
linear in $\theta^{\m\n}$ vanishes since it must be proportional
to $\theta_{\m\n}g_{\m\n}$, $g_{\m\n}$ being the space-time
metric. Hence,
\begin{equation*}
\frac{\langle vac ,n=1|vac,n=0\rangle^{(\theta definst)} }{\langle
vac ,n=0|vac,n=0\rangle^{\phantom{(\theta definst)}}}\thickapprox
\int\frac{d\rho\,d^4x_0 }{\rho^5}\,d_0^{(n_f=1)}(\rho)\,m\rho,
\end{equation*}
 where $d_0^{(n_f=1)}(\rho)$ is the ordinary  ``reduced'' instanton
density~\cite{Vainshtein:1981wh}:
 \begin{equation*}
d_0^{(n_f)}(\rho)=C^{(n_f)}\,\bigg(\frac{8\pi^2}{g^2(\rho)}\bigg)^6\,e^{-\big(\frac{8\pi^2}{g^2(\rho)}\big)}.
\end{equation*}
$C_{n_f}$ is a constant which depends on the number of light
flavours and the regularization scheme.

Unlike $\langle vac ,n=1|vac,n=0\rangle/\langle vac
,n=0|vac,n=0\rangle$, $\langle \psi(x)\psib(y)\rangle^{(\theta
definst)} /\langle vac ,n=0|vac,n=0\rangle$ receives contributions
which are linear in $\theta^{\m\n}$. Indeed, $m\r \ll 1$ leads to
\begin{equation*}
\begin{array}{l}
{\frac{\langle \psi(x)\psi^{\dagger}(y)\rangle^{(\theta definst)}}
{\langle vac ,n=0|vac,n=0\rangle}\thickapprox \int\frac{d\rho d^4x_0
dU}{\rho^5}\,d_0^{(n_f=1)}(\rho)\,m\rho\,
\frac{\psi_0(x-x_0)\psi^\dagger_0(y-x_0)}{-im}}\\
{\phantom{\frac{\langle \psi(x)\psi^{\dagger}(y)\rangle^{(\theta definst)}}
{\langle vac ,n=0|vac,n=0\rangle}}=
\int\frac{d\rho d^4x_0
dU}{-i\rho^4}\,d_0^{(n_f=1)}(\rho)
\Big[\psi^{(0)}_0(x\!-\!x_0)\psi^{(0)\,\dagger}_0(y\!-\!x_0)\!+
h\psi^{(0)}_0(x\!-\!x_0)\psi^{(1a)\,\dagger}_0(y\!-\!x_0)}\\
{\phantom{\frac{\langle \psi(x)\psi^{\dagger}(y)\rangle^{(\theta definst)}}
{\langle vac ,n=0|vac,n=0\rangle}=\int\frac{d\rho }{-i\rho^4}\,d_0^{(n_f=1)}(\,)
}\!+h\psi^{(1a)}_0(x\!-\!x_0)\psi^{(0)\,\dagger}_0(y\!-\!x_0)
\!+\!h\psi^{(0)}_0(x\!-\!x_0)\psi^{(1b)\,\dagger}_0(y\!-\!x_0)}\\
{\phantom{\frac{\langle \psi(x)\psi^{\dagger}(y)\rangle^{(\theta definst)}}
{\langle vac ,n=0|vac,n=0\rangle}=\int\frac{d\rho }{-i\rho^4}\,d_0^{(n_f=1)}(\,)
}+\!h\psi^{(1b)}_0(x\!-\!x_0)\psi^{(0)\,\dagger}_0(y\!-\!x_0)+O(h^2\theta^2)\Big],}\\
\end{array}
\end{equation*}
where $\psi^{(0)}_0(x)$ is the zero mode of the ordinary Dirac
operator in the ordinary instanton field, and $h\psi^{(1a)}_0(x)$
and $h\psi^{(1b)}_0(x)$ are the corrections of order $h\theta^{\m\n}$ to
$\psi^{(0)}_0(x)$ that make, at first order in $h\theta^{\m\n}$,
$\psi_0(x)=\psi^{(0)}_0(x)+h\psi^{(1a)}_0(x)+h\psi^{(1b)}_0(x)$ the
zero mode of the operator ${\cal K}[a_{\m}^{({\rm gsing})}]$. See
Appendix C, for definitions and further details. Since $SU(3)$ is
compact, and following ref.~\cite{Nowak:1988bh}, the averaging
over $SU(3)$ first-order-in-$\theta$-deformed instanton
orientations is carried out by using the first two $SU(3)$
integrals in eq.~(\ref{intgrupo}) of Appendix E.

Let $\tau^a,\,a=1,2\text{ and }3$ denote the upper-left-hand corner embedding of the $SU(2)$ generators
in the generators of $SU(3)$ in the fundamental representation. We define $\tau^\pm_\mu=(\vec{\tau},\mp i),\,{\tau}_{\mu\nu}=\frac{1}{4i}(\tau^-_\mu\tau^+_\nu-\tau^-_\mu\tau^+_\nu)$. Then, taking into account the definitions in eqs.~(\ref{modocerolef}),
(\ref{modocerocomp}) and (\ref{modocerofunc}) of Appendix C, and using the conventions in eq.~(\ref{alphas}) in Appendix E, one obtains the following expressions,
\begin{equation*}
\begin{array}{l}
\psi^{(0)}_{0\,im}(x)\psi^{(0)\,\dagger}_{0\,jn}(y)=\frac{1}{8}\,\phi(x)\phi(y)[(\xslash-\xslash_0)\g_\mu \g_\nu(\yslash-\xslash_0)P_R\,]_{ij}
[U\tau^-_\mu \tau^+_\nu U^\dagger]_{mn},\\
\psi^{(0)}_{0\,im}(x)\psi^{(1a)\,\dagger}_{0\,jn}(y)=\\
\phantom{\,}=\frac{1}{8}\,\phi(x)(\Gamma_{\r\s}(y)(y-x_0)_\r(y-x_0)_\a+\Lambda_{\a\s}(y))[(\xslash-\xslash_0)
\g_\mu \g_\nu (\yslash-\xslash_0) P_R\,]_{ij}[U\tau^-_\mu \tau^+_\nu {\tau}_{\s\a} U^\dagger]_{mn},\\
\psi^{(0)}_{0\,im}(x)\psi^{(1b)\,\dagger}_{0\,jn}(y)=\frac{1}{8}\,\phi(x)\chi^*_{\a\s}(y)(y-x_0)_\alpha[(\xslash-\xslash_0)
\g_\mu \g_\nu \g_\s P_R\,]_{ij}[U\tau^-_\mu \tau^+_\nu U^\dagger]_{mn},\\
\end{array}
\end{equation*}
from which one concludes that the effective Lagrangian ${\cal
L}_{eff}^{(n_f=1)}(x)$ is given by the following equations:
\begin{equation}
\begin{array}{l}
 {\idx\,{\cal L}_{eff}^{(n_f=1)}(x)=\idx\,{\mathfrak L}_0(x)+\idx\,{\mathfrak
 L}_2(x),}\\
{{\mathfrak L}_0(x)=\int
\frac{d\rho}{\r^5}\,d_{0}^{(n_f=1)}(\rho)\,m\r,\; {\mathfrak
 L}_2(x)=\int \frac{d\rho}{\r^5}\,d_{0}^{(n_f=1)}(\rho)\int\frac{d^4p}{(2\pi)^4}\,e^{-ipx}\,{\cal
 Y}_{2}(p),}\\
{{\cal Y}_{2}(p)={\cal Y}^{(0)}_{2}(p)\,+\,h\,{\cal Y}^{(1)}_{2}(p),}\\
{{\cal Y}^{(0)}_{2}(p)=\frac{i}{3}\,\rho\, \overline{\Qc}_R(p) \Qc_L(p),}\\
{{\cal
Y}^{(1)}_{2}(p)=\frac{i}{3}\,\rho\big[\overline{\Sc}_{R}(p)\Qc_L(p)
-\overline{\Qc}_R(p)\Sc_{L}(p)-\overline{\Rc}_{\a\s,R}(p)\g_{\s\a}\Qc_L(p)
-\overline{\Qc}_R(p)\g_{\s\a}\Rc_{\a\s,L}(p)\big],}
\end{array}
\label{oneflavourlag}
\end{equation}
where
\begin{equation*}
\begin{array}{l}
    {\Qc(p)\equiv\phi'(u)u\,\psi(p),  \; \overline{\Qc}(p)\equiv\phi'(u)u\,\psib(p),\;
    \Rc_{\alpha\sigma}(p)\equiv(-\partial_{\r\a}\Gamma'_{\r\s}(u)+\Lambda'_{\a\s}(u))u\,\psi(p),}\\
{\overline{\Rc}_{\alpha\sigma}(p)\equiv(-\partial_{\r\a}\Gamma'_{\r\s}(u)+\Lambda'_{\a\s}(u))u\,\psib(p),\;
    \Sc(p)\equiv\frac{\chi'_{\a\s}(u)p_\a}{u}\g_\s\pslash\,\psi(p),\;
    \overline{\Sc}(p)\equiv\frac{\chi'_{\a\s}(u)p_\a}{u}\,\psib(p)\pslash\g_\s,}\\
\end{array}
\end{equation*}
with $u=\sqrt{p^2}$. Derivatives with respect to $u$ are denoted
by the super-script $'$. In eq.~(\ref{fourier}) in Appendix E, the
functions $\phi(u)$, $\Gamma_{\r\s}(u)$, $\Lambda_{\a\s}(u)$ and
$\chi_{\a\s}(u)$ are given in terms of modified Bessel functions.
${\cal Y}^{(0)}_{2}(p)$ is the ordinary result --see
refs.~\cite{Schafer:1996wv, Diakonov:2002fq}-- and ${\cal
Y}^{(1)}_{2}(p)$ is the first order noncommutative correction.
Note that neither ${\cal Y}^{(0)}_{2}(p)$ nor ${\cal
Y}^{(1)}_{2}(p)$ are invariant under chiral transformations. This
shows that the classical chiral symmetry of the massless theory is
broken in the quantum theory.

One expects that the previous effective Lagrangian gives right
Physics in the low energy regime: $p\rho\ll1$,
$h\theta^{\mu\n}\rho^{-2}\ll1$. Using the low-momentum
approximations in Appendix E, one obtains the following low-energy
expressions for the two-field contribution to ${\cal
L}_{eff}^{(n_f=1)}$:
\begin{equation*}
{\mathfrak L}_{2}(p)=i\int
\frac{d\rho}{\r^5}\,d_{0}^{(n_f=1)}(\rho)\,\frac{4\pi^2\rho^3}{3}\overline{\psi}_R(p)\big[1+h
{\cal
 T}\big]\psi_L(p).
\end{equation*}
Here, ${\cal
T}=-\frac{4}{3p^2\rho^2}(\theta-\tilde{\theta})_{\m\nu}\g_{\a\nu}p_\m
p_\a$, with $\g_{\a\nu}=\frac{1}{4i}[\g_\a,\g_\n]$. Notice that
the ordinary contribution to ${\cal L}_{2}(p)$ just above acts
like a mass term. This interpretation is spoiled by the first
order corrections in $h\theta^{\m\n}$.

\subsection{Two light flavours}

Now, in eq.~(\ref{fermact}), $n_f=2$ and $m_f \r \ll1 $, $f=1,2$.
The effective Lagrangian, ${\cal L}_{eff}^{(n_f=2)}(x)$, that yields
the $m_f\rightarrow 0$ leading contributions to
\begin{equation*}
\frac{\langle vac ,n=1|vac,n=0\rangle}{\langle vac
,n=0|vac,n=0\rangle},\, \frac{\langle
\psi_f(x)\psi^{\dagger}_f(y)\rangle^{(\theta definst)}}{\langle
vac ,n=0|vac,n=0\rangle}\;\,\text{and}\;\,
\frac{\langle\prod_{f=1,2}
\psi_f(x_f)\psi^{\dagger}_f(y_f)\rangle^{(\theta definst)}}
{\langle vac ,n=0|vac,n=0\rangle\phantom{(\theta def)}},
\end{equation*}
reads
\begin{equation}
\begin{array}{l}
{\idx\,{\cal L}_{eff}^{n_f=2}(x)=\idx\,\int
\frac{d\rho}{\r^5}\,d_{0}^{(n_f=2)}(\rho)\prod_{f=1,2}\,[m_f\r+{\cal
Y}_2(x,\psi_f)]}\\
{\phantom{\idx\,{\cal L}_{eff}^{n_f=2}(x)}+ \int
\prod_{j=1,2}\frac{dp_j}{(2\pi)^4}\frac{dq_j}{(2\pi)^4}\delta\big(\sum_{j=1,2}
p_j-\sum_{j=1,2} q_j\big) {\mathfrak L}_4(p_1,p_2,q_1,q_2)}.
\end{array}
\label{lagtwofla}
\end{equation}
Here, ${\cal Y}_2(x,\psi_f)$ is obtained from the corresponding
expression in eq.~(\ref{oneflavourlag}) by performing the Fourier
transform and then applying it to the fermion $\psi_f$ with light
mass $m_f$, and ${\mathfrak L}_4(p_1,p_2,q_1,q_2)$ is given by the
following identities:
\begin{equation}
\begin{array}{l}
    {\mathfrak L}_4(p_1,p_2,q_1,q_2)={\mathfrak L}_4^{(0)}(p_1,p_2,q_1,q_2)+h{\mathfrak L}_4^{(1)}(p_1,p_2,q_1,q_2),\\
    {\mathfrak L}^{(0)}_4(p_1,p_2,q_1,q_2)=\int\,\frac{d\rho}{\r^5}\,d_{0}^{(n_f=2)}(\rho)\,{\cal
    Y}^{(0)}_4(p_1,p_2,q_1,q_2),\\
    {\mathfrak L}^{(1)}_4(p_1,p_2;q_1,q_2)=\int\,\frac{d\rho}{\r^5}\,d_{0}^{(n_f=2)}(\rho)\,{\cal Y}^{(1)}_4(p_1,p_2,q_1,q_2),\\
    {\cal Y}^{(0)}_4(p_1,p_2,q_1,q_2)=\\
    \frac{-\rho^2}{32}\left\{
    \frac{1}{3}(\overline{\Qc}^1_R(p_1)\lambda^a\Qc^1_L(q_1))(\overline{\Qc}^2_R(p_2)
    \lambda^a\Qc^2_L(q_2))+(\overline{\Qc}^1_R(p_1)\g_{\mu\nu}\lambda^a\Qc^1_L(q_1))(\overline{\Qc}^2_R(p_2)\g_{\mu\nu}\lambda^a\Qc^2_L(q_2))\right\},\\
    {\cal Y}^{(1)}_4(p_1,p_2,q_1,q_2)=\\
    \frac{-\rho^2}{32}\left\{\frac{1}{3}(\overline{\Sc}^1_{R}(p_1)\lambda^a\Qc^1_L(q_1)-\overline{\Qc}^1_R(p_1)
    \lambda^a \Sc^1_{L}(q_1))(\overline{\Qc}^2_R(p_2)\lambda^a\Qc^2_L(q_2))\right.\\
    \phantom{\Delta{\mathfrak L}^{(h)}}\left.+(\overline{\Sc}^1_{R}(p_1)\g_{\mu\nu}\lambda^a\Qc^1_L(q_1)-
    \overline{\Qc}^1_R(p_1)\g_{\mu\nu}\lambda^a \Sc^1_{L}(q_1))(\overline{\Qc}^2_R(p_2)\g_{\mu\nu}\lambda^a\Qc^2_L(q_2))\right\}\\
    +\frac{\rho^2}{32}\left\{\frac{1}{3}(\overline{\Qc}^1_{R}(p_1)\g_{\s\a}\lambda^a\Rc^1_{\a\s,L}(q_1)+
    \overline{\Rc}^1_{\a\s,R}(p_1)\g_{\s\a}\lambda^a \Qc^1_{L}(q_1))(\overline{\Qc}^2_R(p_2)\lambda^a\Qc^2_L(q_2))\right.\\
    \phantom{\Delta{\mathfrak L}^{(h)}}+(\overline{\Qc}^1_R(p_1)\lambda^a\Rc^1_{\a\s,L}(q_1)+
    \overline{\Rc}^1_{\a\s,R}(p_1)\lambda^a\Qc^1_L (q_1))(\overline{\Qc}^2_R(p_2)\g_{\s\a}\lambda^a\Qc^2_L(q_2))\\
    \phantom{{\mathfrak L}}
    \left.+i\big[(\overline{\Qc}^1_R(p_1)\g_{\a\nu}\lambda^a\Rc^1_{\a\s,L}(q_1)-
    \overline{\Rc}^1_{\a\s,R}(p_1)\g_{\a\nu}\lambda^a\Qc^1_L (q_1))(\overline{\Qc}^2_R(p_2)\g_{\nu\s}\lambda^a\Qc^2_L(q_2))-(\a\leftrightarrow\s)\big]\right\}\\
    +(1\leftrightarrow2).
\end{array}
\label{fourflavourlag}
\end{equation}
$\Qc^f(p)$, $\overline{\Qc}^f(p)$, $\Rc^f_{\alpha\sigma}(p)$,
$\overline{\Rc}^f_{\alpha\sigma}(p)$,  $\Sc^f(p)$ and
$\overline{\Sc}^f(p)$ in the previous equation are defined by the
following equations:
\begin{equation*}
\begin{array}{l}
   { \Qc^f(p)\equiv\phi'(u)u\,\psi_f(p),
   \;\overline{\Qc}^f(p)\equiv\phi'(u)u\,\psib_f(p),\;
 \Rc^f_{\alpha\sigma}(p)\equiv(-\partial_{\r\a}\Gamma'_{\r\s}(u)+\Lambda'_{\a\s}(u))u\,\psi_f(p),}\\
{\overline{\Rc}^f_{\alpha\sigma}(p)\equiv(-\partial_{\r\a}\Gamma'_{\r\s}(u)+\Lambda'_{\a\s}(u))u\,\psib_f(p),
    \;
    \Sc^f(p)\equiv\frac{\chi'_{\a\s}(u)p_\a}{u}\g_\s\pslash\,\psi_f(p),}\\
{\overline{\Sc}^f(p)\equiv\frac{\chi'_{\a\s}(u)p_\a}{u}\,\psib_f(p)\pslash\g_\s,}\\
\end{array}
\end{equation*}
where $u=\sqrt{p^2}$ and  the derivatives with respect to $u$ are
denoted by the super-script $'$. The functions $\phi(u)$,
$\Gamma_{\r\s}(u)$, $\Lambda_{\a\s}(u)$ and $\chi_{\a\s}(u)$ are
given in terms of modified Bessel functions in eq.~(\ref{fourier})
of Appendix E.

To obtain ${\cal L}_{eff}^{(n_f=2)}(x)$ in eq.~(\ref{lagtwofla}),
averages over $SU(3)$ are to be carried out with the help of
the first three equalities in eq.~(\ref{intgrupo}) of Appendix E.
Notice that if we set $h=0$ in eq.~(\ref{lagtwofla}) the ordinary
't Hooft vertex for two light flavours~\cite{Schafer:1996wv,
Diakonov:2002fq} is recovered.

As in the one-flavour case, we expect that ${\cal
L}_{eff}^{(n_f=2)}(x)$ in eq.~(\ref{lagtwofla}) gives the correct
Physics in the low momenta limit: $p_i\r\ll1, q_i\r\ll1$, $i=1,2$.
In this limit,  ${\mathfrak L}_4(p_1,p_2,q_1,q_2)$ boils down to
\begin{equation*}
\begin{array}{l}
{{\mathfrak
L}_4(p_1,p_2,q_1,q_2)\approx -\int\,\frac{d\rho}{\r^5}\,d_{0}^{(n_f=2)}(\rho)\,\Big[\frac{3}{32}\big(\frac{4\pi^2\rho^3}{3}\big)^2\big\{
(\overline{\psi}_{1\,R}(p_1)\lambda^a\psi_{1\,L}(q_1))
(\overline{\psi}_{2\,R}(p_2)\lambda^a\psi_{2\,L}(q_2))}\\
{\quad\quad\;\,\phantom{{\mathfrak
L}_4(p_1,p_2,q_1,q_2)=\int\,\frac{d\rho}{\r^5}\,d_{0}^{(n_f=2}(\rho)\,}+3(\overline{\psi}_{1\,R}(p_1)\g_{\mu\nu}\lambda^a\psi_{1\,L}(q_1))
(\overline{\psi}_{2\,R}(p_2)\g_{\mu\nu}\lambda^a\psi_{2\,L}(q_2))\big\}\Big]}\\
{\phantom{{\mathfrak L}_4}
-\,h\,\int\,\frac{d\rho}{\r^5}\,d_{0}^{(n_f=2)}(\rho)\,\Big[\frac{3}{32}\left(\frac{4\pi^2\rho^3}{3}\right)^2\big\{(\overline{\psi}_{1\,R}(p_1)
[\overline{{\cal O}}(p_1)-{\cal O}(q_1)]
\lambda^a\psi_{1\,L}(q_1))(\overline{\psi}_{2\,R}(p_2)\lambda^a\psi_{2\,L}(q_2))}\\
{\phantom{{\mathfrak L}_4(p_1,p_2,q_1,q_2) +\int\,}+
3(\overline{\psi}_{1\,R}(p_1)[\overline{\cal O}(p_1)
\g_{\mu\nu}-\g_{\mu\nu}{\cal
O}(q_1)]\lambda^a\psi_{1\,L}(q_1))(\overline{\psi}_{2\,R}(p_2)\g_{\mu\nu}\lambda^a\psi_{2\,L}(q_2))\big\}}\\
{\quad\;+(1\leftrightarrow2)\Big],}\\
\end{array}
\label{lowmomentatwoflav}
\end{equation*}
where
\begin{equation}
\overline{{\cal O}}(p)=\frac{i}{3\r^2
p^2}\,(\theta-\tilde{\theta})_{\m\n}\pslash p_\m \g_\n\quad
\text{and}\quad {\cal O}(p)=\frac{i}{3\r^2
p^2}\,(\theta-\tilde{\theta})_{\m\n}\g_\n\pslash p_\m .
\label{calO}
\end{equation}
To obtain eq.~(\ref{lowmomentatwoflav}), we have used the
approximations in eq.~(\ref{asint}) of Appendix E. Of course, if
the contributions linear in $\theta^{\m\n}$ are dropped one
obtains the ordinary contributions in ref.~\cite{Shifman:1979uw}.

\subsection{Three light flavours}

We shall finally give the effective Lagrangian, ${\cal
L}_{eff}^{(n_f=3)}(x)$, for the case of three light fermions: the
fermionic action is the action in eq.~(\ref{fermact}) for $n_f=3$.

Let us first introduce some notation. ${\cal
Y}_4(x,\psi_f,\psi_{f'})$ is defined by
\begin{equation*}
{\cal Y}_4(x,\psi_f,\psi_{f'})=\int
\prod_{j=1}^2\frac{dp_j}{(2\pi)^4}\frac{dq_j}{(2\pi)^4}\,e^{-i(\sum_{j=1}^2
p_i-\sum_{j=1}^2 q_i)x}\big[{\cal
Y}^{(0)}_4(p_1,p_2,q_1,q_2)+h{\cal
Y}^{(1)}_4(p_1,p_2,q_1,q_2)\big],
\end{equation*}
once $\psi_1$ and $\psi_2$  are replaced with $\psi_f$ and
$\psi_{f'}$, respectively, both in ${\cal
Y}^{(0)}_4(p_1,p_2,q_1,q_2)$ and ${\cal
Y}^{(1)}_4(p_1,p_2,q_1,q_2)$, which are given in
eq.~(\ref{fourflavourlag}). ${\cal Y}_2(x,\psi_f)$ shall denote
the same quantity that occurred in eq.~(\ref{lagtwofla}).
 Then, ${\cal L}_{eff}^{(n_f=3)}(x)$ is given by the expressions that follow
\begin{equation*}
\begin{array}{l}
{\idx\, {\cal L}_{eff}^{(n_f=3)}(x)=\idx\,\int
\frac{d\rho}{\r^5}\,d_{0}^{(n_f=3)}(\rho)\prod_{f=1,2,3}\,[m_f\r+{\cal
Y}_2(x,\psi_f)]}\\
{\phantom{\;}+\idx\,\int
\frac{d\rho}{\r^5}\,d_{0}^{(n_f=3)}(\rho)\Big\{[m_1\r+{\cal
Y}_2(x,\psi_1)]{\cal
Y}_4(x,\psi_2,\psi_3)+[m_2\r+{\cal
Y}_2(x,\psi_2)]{\cal
Y}_4(x,\psi_1,\psi_3)}\\
{\phantom{\;}+[m_3\r+{\cal
Y}_2(x,\psi_3)]{\cal
Y}_4(x,\psi_1,\psi_2)\Big\}
}\\
{\phantom{\;}+\int
\prod_{j=1}^3\frac{dp_j}{(2\pi)^4}\frac{dq_j}{(2\pi)^4}\delta\big(\sum_{j=1}^3
p_j-\sum_{j=1}^3 q_j\big)\; {\mathfrak
L}_6(p_1,p_2,p_3,q_1,q_2,q_3),}\\
\end{array}
\end{equation*}
where
\begin{equation*}
\begin{array}{l}
    {\mathfrak L}_6(p_1,p_2,p_3,q_1,q_2,q_3)={\mathfrak L}_6^{(0)}(p_1,p_2,p_3,q_1,q_2,q_3)+
h{\mathfrak L}_6^{(1)}(p_1,p_2,p_3,q_1,q_2,q_3),\\
    {\mathfrak L}_6^{(0)}(p_1,p_2,p_3,q_1,q_2,q_3)=\int \frac{d\rho}{\r^5}\,d_{0}^{(n_f=3)}(\rho)\,\bigg[\\
    \frac{-i\rho^3}{64}\left\{d_{abc}\big[-\frac{1}{15}(\overline{\Qc}_R^1(p_1)
\lambda^a\Qc_L^1(q_1))(\overline{\Qc}_R^2(p_2)
    \lambda^b\Qc_L^2(q_2))(\overline{\Qc}_R^3(p_3)\lambda^c\Qc_L^3(q_3))\right.\\
    +\left(\frac{1}{5}(\overline{\Qc}_R^1(p_1)\g_{\mu\nu}\lambda^a\Qc_L^1(q_1))
(\overline{\Qc}_R^2(p_2)\g_{\mu\nu}
    \lambda^b\Qc_L^2(q_2))(\overline{\Qc}_R^3(p_3)\lambda^c\Qc_L^3(q_3))+
(3\leftrightarrow1)+(3\leftrightarrow2)\right)\big]\\
\left.-\frac{2}{3}f_{abc}(\overline{\Qc}_R^1(p_1)\g_{\a\b}\lambda^a\Qc_L^1(q_1))
    (\overline{\Qc}_R^2(p_2)\g_{\b
    \delta}\lambda^b\Qc_L^2(q_2))(\overline{\Qc}_R^3(p_3)\g_{\delta\a}\lambda^c\Qc_L^3(q_3))\right\}\bigg],\\
{\mathfrak L}_6^{(1)}(p_1,p_2,p_3,q_1,q_2,q_3)=\int \frac{d\rho}{\r^5}\,d_{0}^{(n_f=3)}(\rho)\,\bigg[\\
\frac{-i\rho^3}{64}\left\{d_{abc}\big[-\frac{1}{15}(\overline{\Sc}_{R}^1(p_1)\lambda^a\Qc_L^1(q_1)-
\overline{\Qc}^1_R(p_1)
\lambda^a\Sc_L^1(q_1))(\overline{\Qc}_R^2(p_2)\lambda^b\Qc_L^2(q_2))
(\overline{\Qc}_R^3(p_3)\lambda^c\Qc_L^3(q_3))\right.\\
    \left.+\frac{1}{5}\left((\overline{\Sc}_{R}^1(p_1)\g_{\mu\nu}\lambda^a\Qc_L^1(q_1)\!-\!\overline{\Qc}^1_R(p_1)\g_{\mu\nu}
    \lambda^a\Sc^1_L(q_1))(\overline{\Qc}_R^2(p_2)\g_{\mu\nu}\lambda^b\Qc_L^2(q_2))
(\overline{\Qc}_R^3(p_3)\lambda^c\Qc_L^3(q_3))\right.\right.\\
    \!+\!(3\leftrightarrow2)+(\overline{\Sc}^1_R(p_1)\lambda^a\Qc_L^1(q_1)\!-\!\overline{\Qc}_R^1(p_1)\lambda^a\Sc_L^1(q_1))
    (\overline{\Qc}_R^2(p_2)\g_{\mu\nu}\lambda^b\Qc_L^2(q_2))
(\overline{\Qc}_R^3(p_3)\g_{\mu\nu}\lambda^c\Qc_L^3(q_3))\big)\big]\\
    \!-\!\frac{2}{3}f_{abc}(\overline{\Sc}_R^1(p_1)\g_{\a\b}\lambda^a\Qc_L^1(q_1)\!-\!\overline{\Qc}_R^1(p_1)\g_{\a\b}
    \lambda^a\Sc_L^1(q_1))(\overline{\Qc}_R^2(p_2)\g_{\b \delta}\lambda^b\Qc_L^2(q_2))
(\overline{\Qc}_R^3(p_3)\g_{\delta\a}\lambda^c\Qc_L^3(q_3))\\
    \big\}
    -\frac{i\rho^3}{128}\big\{\frac{2}{5}d_{abc}\big[\\
    \frac{1}{3}(\overline{\Qc}^1_R(p_1)\g_{\s\a}\lambda^a\Rc^1_{\a\s,L}(q_1)+\overline{\Rc}^1_{\a\s,R}(p_1)
    \g_{\s\a}\lambda^a\Qc_L^1(q_1))(\overline{\Qc}^2_R(p_2)\lambda^b\Qc^2_L(q_2))
(\overline{\Qc}^3_R(p_3)\lambda^c\Qc^3_L(q_3))\\
    -\big((\overline{\Qc}^1_R(p_1)\lambda^a\Rc^1_{\a\s,L}(q_1)+\overline{\Rc}^1_{\a\s,R}(p_1)\lambda^a\Qc_L^1(q_1))(
    \overline{\Qc}^2_R(p_2)\lambda^b\Qc^2_L(q_2))(\overline{\Qc}^3_R(p_3)\g_{\s\a}\lambda^c\Qc^3_L(q_3))\\
    +(2\leftrightarrow3)\big)
    -(\overline{\Qc}^1_R(p_1)\g_{\s\a}\lambda^a\Rc^1_{\a\s,L}(q_1)+\overline{\Rc}^1_{\a\s,R}(p_1)\g_{\s\a}
    \lambda^a\Qc_L^1(q_1))(\overline{\Qc}^2_R(p_2)\g_{\mu\nu}\lambda^b\Qc^2_L(q_2))\cdot\\
    \cdot(\overline{\Qc}^3_R(p_3)\g_{\mu\nu}\lambda^c\Qc^3_L(q_3))
    +i\big( (\overline{\Qc}^1_R(p_1)\g_{\s\nu}\lambda^a\Rc^1_{\a\s,L}(q_1)-
\overline{\Rc}^1_{\a\s,R}(p_1)\g_{\s\nu}\lambda^a\Qc_L^1(q_1))\cdot\\
    \cdot(\overline{\Qc}^2_R(p_2)\g_{\nu\a}\lambda^b\Qc^2_L(q_2))(\overline{\Qc}^3_R(p_3)
    \lambda^c\Qc^3_L(q_3)) +(2\leftrightarrow3)-(\alpha\leftrightarrow\s)\big)
    \big]+\frac{1}{3}f_{abc}\big[\\
    \big((\overline{\Qc}^1_R(p_1)\lambda^a\Rc^1_{\a\s,L}(q_1)+\overline{\Rc}^1_{\a\s,R}(p_1)
    \lambda^a\Qc_L^1(q_1))(\overline{\Qc}^2_R(p_2)\g_{\a\nu}\lambda^b\Qc^2_L(q_2))
(\overline{\Qc}^3_R(p_3)\g_{\nu\s}\lambda^c\Qc^3_L(q_3))\\
    -(\alpha\leftrightarrow\s)\big)-2i\big((\overline{\Qc}^1_R(p_1)\g_{\mu\nu}\lambda^a\Rc^1_{\a\s,L}(q_1)
-\overline{\Rc}^1_{\a\s,R}(p_1)
    \g_{\mu\nu}\lambda^a\Qc_L^1(q_1))(\overline{\Qc}^2_R(p_2)\g_{\mu\nu}\lambda^b\Qc^2_L(q_2))\\
    (\overline{\Qc}^3_R(p_3)\g_{\s\a}\lambda^c\Qc^3_L(q_3))-(2\leftrightarrow3)\big)\big]
    \big\}+(1\leftrightarrow2)+(1\leftrightarrow 3)\bigg].
\end{array}
\end{equation*}
To obtain the average over the $SU(3)$ orientations leading to the
previous equation, we have employed the last integral in
eq.~(\ref{intgrupo}) of Appendix E; some values of the structure constants of $SU(3)$ were substituted. The low momenta --i.e., $p_i\rho\ll1, q_i\rho\ll1$,
 $i=1,2,3$--  approximation to  ${\mathcal L}_6(p_1,p_2,p_3,q_1,q_2,q_3)$ can be worked out with the help of eq.~(\ref{asint}) in Appendix E. We display the value
of ${\mathcal L}_6(p_1,p_2,p_3,q_1,q_2,q_3)$ in this low momenta limit in the following equation:
\begin{equation*}
\begin{array}{l}
{\mathfrak L}_6(p_1,p_2,p_3,q_1,q_2,q_3)\approx\int \frac{d\rho}{\r^5}\,d_{0}^{(n_f=3)}(\rho)\big(\frac{-i3^3}{64}\big)\big(\frac{4\pi^2\rho^3}{3}\big)^3\,\bigg[\\
d_{abc}\big[-\frac{1}{15}
(\overline{\psi}_{1\,R}(p_1)\lambda^a\psi_{1\,L}(q_1))(\overline{\psi}_{2\,R}(p_2)\lambda^b\psi_{2\,L}(q_2))
(\overline{\psi}_{3\,R}(p_3)\lambda^c\psi_{3\,L}(q_3))\\
+\frac{1}{5}((\overline{\psi}_{1\,R}(p_1)\g_{\mu\nu}\lambda^a\psi_{1\,L}(q_1))
(\overline{\psi}_{2\,R}(p_2)\g_{\mu\nu}\lambda^b\psi_{2\,L}(q_2))(\overline{\psi}_{3\,R}(p_3)\lambda^c\psi_{3\,L}(q_3))+
(1\leftrightarrow 3)+(2\leftrightarrow 3))\big]\\
\phantom{\Delta{\mathfrak L}}-\frac{2}{3}f_{abc}
(\overline{\psi}_{1\,R}(p_1)\g_{\alpha\beta}\lambda^a\psi_{1\,L}(q_1))(\overline{\psi}_{2\,R}(p_2)
\g_{\b\delta}\lambda^b
\psi_{2\,L}(q_2))
(\overline{\psi}_{3\,R}(p_3)\g_{\delta\a}\lambda^c\psi_{3\,L}(q_3))\\
+h\big\{d_{abc}\big[-\frac{1}{15}
(\overline{\psi}_{1\,R}(p_1)[\overline{\cal O}(p_1)-{\cal O}(q_1)]\lambda^a\psi_{1\,L}(q_1))
(\overline{\psi}_{2\,R}(p_2)\lambda^b\psi_{2\,L}(q_2))(\overline{\psi}_{3\,R}(p_3)\lambda^c\psi_{3\,L}(q_3))\\
+\frac{1}{5}((\overline{\psi}_{1\,R}(p_1)[\overline{\cal O}(p_1)
\g_{\mu\nu}-\g_{\mu\nu}{\cal O}(q_1)]\lambda^a\psi_{1\,L}(q_1))(\overline{\psi}_{2\,R}(p_2)\g_{\mu\nu}\lambda^b\psi_{2\,L}(q_2))
(\overline{\psi}_{3\,R}(p_3)\lambda^c\psi_{3\,L}(q_3))\\
+(2\leftrightarrow3)\\
+(\overline{\psi}_{1\,R}(p_1)[\overline{\cal O}(p_1)-{\cal O}(q_1)]
\lambda^a\psi_{1\,L}(q_1))(\overline{\psi}_{2\,R}(p_2)\g_{\mu\nu}\lambda^b\psi_{2\,L}(q_2))(\overline{\psi}_{3\,R}(p_3)\g_{\mu\nu}
\lambda^c\psi_{3\,L}(q_3)))\big]-\frac{2}{3}\\
f_{abc}(\overline{\psi}_{1\,R}(p_1)[\overline{\cal O}(p_1)
\g_{\alpha\beta}-\g_{\alpha\beta}{\cal O}(q_1)]\lambda^a\psi_{1\,L}(q_1))
(\overline{\psi}_{2\,R}(p_2)\g_{\b\delta}
\lambda^b\psi_{2\,L}(q_2))
(\overline{\psi}_{3\,R}(p_3)\g_{\delta\a}\lambda^c
\psi_{3\,L}(q_3))\\
+(1\leftrightarrow2)+(1\leftrightarrow3)\big\}\bigg].\\
\end{array}
\end{equation*}
$\overline{\cal O}(p)$ and ${\cal O}(p)$ are defined in eq.~(\ref{calO}).

\section{Summary and outlook}

In the main body of this  paper,  we have obtained the   following results
for noncommutative $SU(3)$ gauge theories with one, two and three light 
Dirac fermionic flavours, when (Euclidean) time is commutative:
\begin{enumerate}[i.]
\item There are no solutions at any order in the formal power
expansion in $h\theta^{\m\n}$ to the noncommutative
(anti-)self-duality equations. This result holds for $SU(N)$ as
well.

\item At first order in $h\theta^{\m\n}$, ordinary instantons can
be given a $\theta^{\m\n}$-dependent piece so that the resulting
field configuration satisfies the noncommutative Yang-Mills
equations. This field configuration --that we have called
first-order-in-$\theta$-deformed instanton-- has  
instanton number equal to one, and interpolates,
along Euclidean time, between vacua that differ in one unit of the
winding number. We have also computed the most general
first-order-in-$\theta$-deformed instanton.

\item We have shown that in the first-order-in-$\theta$-deformed instanton
transition a coupling between light left handed and right handed fermions is produced,
thus showing that the classical $U(1)_A$ symmetry of the massless theory is broken
at the quantum level. We have computed
the 't Hooft vertices that describe this coupling and seen that they receive
contributions that are linear in $h\theta^{\m\n}$, these contributions being nonlocal even in the low momenta limit.

\end{enumerate}

In the Appendices --see Appendices A, B, and C, respectively--, we have further obtained the following results:

\begin{enumerate}[i.]

\item That the self-duality equations for noncommutative $SU(N)$ Yang-Mills theory
have solutions that are formal power series in $h\theta^{\m\n}$ if, and only
if, $\theta^{\m\n}$ is self-dual. These solutions are the ordinary instantons
and multi-instantons. Analogously, the noncommutative anti-self-duality equations for $SU(N)$ have solutions that are formal power series in $h\theta^{\m\n}$
if, and only if, $\theta^{\m\n}$ is anti-self-dual. The solutions in question are the ordinary anti-instantons and anti-multi-instantons.

\item That all the noncommutative classical vacua of the noncommutative $SU(N)$ Yang-Mills theory that are formal power series in $h\theta^{\m\n}$ are given by  the Seiberg-Witten transform of some ordinary vacua $ig(\vec{x})\partial_{i}g(\vec{x})^{\dagger}$, where $g(\vec{x})$ is an
ordinary gauge transformation. And

\item that the corrections to the zero mode of the
$\theta$-deformed Dirac operator can be worked out explicitly --this we have done-- at first
order in $h\theta^{\m\n}$ for an arbitrary
first-order-in-$\theta$-deformed instanton.

\end{enumerate}

The analysis and computations carried out in this paper should be
extended at least in two directions. On the one hand, it would be very 
interesting to see whether, for commutative time, there are topologically nontrivial solutions to the
noncommutative Euclidean classical equations of motion  that are
not formal power series in $h\theta^{\m\n}$. This is a highly
nontrivial issue since the action of the theory has been defined so far
as a formal power series in  $h\theta^{\m\n}$. Some kind of
re-summation of the power series expansion would thus be needed,
or perhaps one should define the Seiberg-Witten by expanding it in terms
of a different object~\cite{Jurco:2001rq}. On the other hand, it
will be interesting to work out the second order 
in $h\theta^{\m\n}$ corrections to the instanton density and 't Hooft
vertices  that we have obtained. This is quite an involved 
computation since it will demand the use of the constrained
instanton method~\cite{Affleck:1980mp, Glerfoss:2005mg} or the
valley method~\cite{Balitsky:1986qn, Khoze:1991sa, Aoyama:1991ca}
to carry it out. Indeed, as we show in Appendix D, there are no
topologically nontrivial field configurations that are formal
power series in $h\theta^{\m\n}$ and leave the noncommutative
$SU(N)$ Yang-Mills action stationary at second order in
$h\theta^{\m\n}$. Actually, at second order in $h\theta^{\m\n}$,
the size, $\rho$, of the first-order-in-$\theta$-deformed
instanton does  not yield a zero mode of the quantum bosonic kinetic term
in a background that differs from our
first-order-in-$\theta$-deformed instanton by a term quadratic in
$h\theta^{\m\n}$. Indeed, as can be shown by substituting the
r.h.s. of eq.~(\ref{fordertheta}) in the r.h.s. of
eq.~(\ref{ncymaction}), the noncommutative action acquires a dependence on 
$\rho$ of the type $\rho^{-4}$:
\begin{equation*}
S_{NCYM}=\frac{8\pi^2}{g^2}+\frac{8h^2\,\pi^2}{7g^2\rho^4}(\theta^{\mu\n}-\tilde{\theta}^{\m\n})^2\,+\,O(h^3\theta^3).
\end{equation*}
Notice that if we add to our first-order-in-$\theta$-deformed instanton in eq.~(\ref{fordertheta}) an arbitrary piece that is quadratic in $h\theta^{\m\n}$, the previous value of $S_{NCYM}$ gets no correction at second order in
$h\theta^{\m\n}$.
Now, $\rho$ gives rise to a quasi-zero mode in the sense of ref.~\cite{Balitsky:1986qn}, so that the technique developed in the latter reference can be used to compute higher order corrections in $h\theta^{\m\n}$.

\newpage
\section{Acknowledgments}

This work has been financially supported in part by MEC through grant
BFM2002-00950. The work of C. Tamarit has  also received financial support
from MEC trough FPU grant AP2003-4034. We would like to express our gratitude 
to Professors D. Diakonov and D.K. Nielsen  for instructing us on some 
instanton issues.

\newpage

\section*{Appendix A}
\renewcommand{\theequation}{A.\arabic{equation}}

In this Appendix we shall consider an arbitrary  $\theta^{\m\n}$
and seek for topologically non-trivial solutions, $a_{\m}$, to the
noncommutative (anti-)self-duality equation --eq.~(\ref{sdequation})-- 
that are given by the formal power series in $h\theta^{\mu\nu}$ in
eq.~(\ref{expasionofa}). Throughout this Appendix the noncommutative
gauge field and field strength will be defined in terms of the ordinary
fields by means of the standard Seiberg-Witten map --see
eqs.~(\ref{cswmap}) and~(\ref{swstrengh}). We shall show that for
non-vanishing instanton number the solutions we seek for exist if, 
and only if, $\theta^{\m\n}$ is
(anti-)self-dual and, further, that these solutions are the ordinary (anti-)instantons and 
(anti-)multi-instantons.
In this Appendix $\tilde{v}_{\m\n}$ shall denote the dual
of a given tensor $v_{\m\n}$:
$\tilde{v}_{\m\n}=\frac{1}{2}\epsilon_{\m\n\r\s}v_{\r\s}$.

 The expansion of $a_{\m}[h]$ in eq.~(\ref{expasionofa}) leads to
 the following expansion of its field strength:
 \begin{equation}
f_{\mu\nu}[h]=f^{(0)}_{\mu\nu}+\sum_{l=1}^{\infty}\, h^l f^{(l)}_{\mu\nu},
\label{fmnexpan}
\end{equation}
where $f^{(l)}_{\mu\nu}$ is a homogeneous polynomial in $\theta^{\mu\nu}$ of degree $l$. Substituting  in eq.~(\ref{swstrengh}), both the previous result and eq.~(\ref{expasionofa}), one obtains the following expansion of $F_{\m\n}[a_{\s}[h]]$ in powers of $h\theta^{\mu\nu}$:
\begin{equation}
F_{\mu\nu}[a_\s [h]]=f^{(0)}_{\mu\nu}+
\sum_{l=1}^\infty h^l f^{(l)}_{\mu\nu}+\sum_{l=1,k=0}^\infty h^{l+k}F^{(l,k)}_{\m\n},
\label{Fexpansion2}
\end{equation}
where $F^{(l,k)}_{\m\n}$ is given by
\begin{equation}
F^{(l,k)}_{\m\n}=\frac{1}{l!k!}\frac{d^k}{dt^k}
\frac{d^{l-1}}{d^{l-1}h}\overset{\circ}{F}_{\mu\nu}[a_\s[t]]\Big\vert_{h=t=0}.
\label{Fexpansion3}
\end{equation}
$a_\s [t]$ is obtained from $a_\s [h]$ in eq.~(\ref{expasionofa})
by replacing $h$ with $t$. $\overset{\circ}{F}_{\mu\nu}[a_\s]$ is
equal to $\frac{d}{dh}{F}_{\mu\nu}[a_\s]$ as defined in
eq.~(\ref{firstderiv}).

We shall show below that $F_{\mu\nu}[a_\s [h]]$ as defined above
--see the previous equations-- is self-dual with non-vanishing
Pontrjagin number if, and only if, both $\theta_{\mu\nu}$ and
$f_{\m\n}[h]$ in eq.~(\ref{fmnexpan}) are self-dual. We shall
carry out this proof by induction. At order $h^0$ and $h$, the
self-duality equation for $F_{\mu\nu}[a_\s [h]]$ in
eq.~(\ref{Fexpansion2}) is equivalent to
\begin{equation}
    f^{(0)}_{\mu\nu}=\tilde{f}^{(0)}_{\mu\nu}
    \label{f0}
\end{equation}
and
\begin{equation}
f^{(1)}_{\m\n}+F^{(1,0)}_{\m\n}=\tilde{f}^{(1)}_{\m\n}+\tilde{F}^{(1,0)}_{\m\n},
\label{order1}
\end{equation}
respectively. Since we shall look for solutions with a non-zero
Pontrjagin number, we must demand that $f^{(0)}_{\mu\nu}$  does not vanish
--see eq.~(\ref{swpontri})-- and recall that $f^{(1)}_{\mu\nu}$ does not depend
on $h$. Now, working out the trace over the $SU(N)$ generators on both sides of
eq.~(\ref{order1}) and using eq.~(\ref{f0}), one obtains the following
equation:
\begin{equation}
\sum_a\frac{1}{2}\big[(f^{(0)\,a}_{12})^2+(f^{(0)\,a}_{13})^2+(f^{(0)\,a}_{23})^2\big](\theta-\tilde{\theta})_{\mu\nu}=0,
\label{tet-atet}
\end{equation}
where ``$a$'' is the colour index of the field strength. Now, for a non-zero
$f^{(0)}_{\mu\nu}$, this equation holds if, and only if, $\theta_{\mu\nu}$ is self-dual. For a self-dual $\theta_{\m\n}$ and a self-dual $f^{(0)}_{\mu\nu}$,
eq.~(\ref{order1}) boils down to
\begin{equation}
 f^{(1)}_{\m\n}=\tilde{f}^{(1)}_{\m\n}.
 \label{f1}
\end{equation}

To show that if $\theta_{\mu\n}$ is self-dual, the self-duality equation
for $F_{\m\n}[a_{\s}[h]]$ in eq.~(\ref{Fexpansion2}) is equivalent to
the self-duality equation for $f_{\m\n}$ in eq.~(\ref{fmnexpan}), we shall
prove first --by induction-- the following statement: if $\theta_{\m\n}$, $f^{(0)}_{\mu\nu}$,...,$f^{(k-1)}_{\mu\nu}$ are self-dual, then, so are $F^{(m,0)}$,...,$ F^{(m,k-1)}$ for all $m\geq 1$.

 From eqs.~(\ref{Fexpansion3}) and~(\ref{firstderiv}), one obtains that
\begin{equation*}
F^{(1,l)}_{\mu\nu}=\sum_{m+n=l}\left(\frac{1}{2}\,\theta^{\kappa\lambda}\{f^{(m)}_{\mu\kappa},f^{(n)}_{\nu\lambda}\}-\frac{1}{4}\,\theta^{\kappa\lambda}\{a^{(m)}_\kappa,[(\partial_\lambda+D_\lambda)f_{\mu\nu}]^{(n)}\}\right).
\end{equation*}
    For $l\leq k-1$, the previous expression involves $f^{(m)}_{\m\n},\,m\leq k-1,$ which are self-dual by
hypothesis; this clearly makes the second term in the previous expression self-dual in $\mu,\nu$. The first term is also self-dual in $\mu,\nu$ as a consequence of the self-duality of $\theta_{\m\n}$, the self-duality of $f^{(m) }_{\m\n},\,m\leq k-1,$ and the property
 \begin{equation}
\epsilon_{\mu\nu\a\b}\epsilon_{\g\delta\lambda\b}=\delta_{\mu\g}\delta_{\nu\delta}\delta_{\a\lambda}-(\lambda\leftrightarrow\delta)+\delta_{\mu\delta}\delta_{\nu\lambda}\delta_{\a\gamma}-(\lambda\leftrightarrow\gamma)+\delta_{\mu\lambda}\delta_{\nu\gamma}\delta_{\a\delta}-(\gamma\leftrightarrow\delta).
\label{prodepsilon}
\end{equation}
This proves the previous statement for $m=1$. We shall assume  in
the sequel that the statement holds for  $m\leq n-1$. Now,
eqs.~(\ref{Fexpansion3}) and~(\ref{firstderiv})  lead to
\begin{equation}
\begin{array}{l}
{    F^{(n,j)}_{\mu\nu}=}\\
{\frac{1}{n!j!}\frac{d^{n-1}}{dh^{n-1}}\frac{d^j}{dt^j}\left.\left(
    \frac{1}{2}\theta^{\kappa\lambda}\{F_{\mu\kappa}[a_\s[t]],F_{\nu\lambda}
[a_\s[t]]\}_\star-\frac{1}{4}\theta^{\kappa\lambda}\{A_\kappa[a_\s[t]],(\partial_l+D_l)F_{\mu\nu}[a_\s[t]]\}_\star\right)\right|_{h=t=0}.}
\label{m+1}
\end{array}
\end{equation}
Taking into account the degrees of the derivatives in the previous
expression, one readily realizes that the  $F^{(i,l)}_{\m\n}$'s
that occur on the r.h.s. of this equation have $i\leq n-1, l\leq
j$. These $F^{(i,j)}_{\m\n}$'s are, by hypothesis, self-dual if
$j\leq k-1$. One thus concludes that the second term on the r.h.s
of eq.~(\ref{m+1}) is self-dual in $\mu,\nu$.
Eq.~(\ref{prodepsilon}) and the fact that $\theta_{\m\n}$ and
$F^{(i,l)}_{\m\n}$, $i<n$, $l<k$ are self-dual imply that the
first term on the r.h.s. of eq.~(\ref{m+1}) is also self-dual.
This concludes the proof of the statement made at the beginning of
the paragraph starting right below eq.~(\ref{f1}).

We are now ready to go on with the proof that the only
topologically non-trivial solutions to eq.~(\ref{sdequation}) that are formal
power series in $h\theta^{\m\n}$ are the ordinary instantons and multi-instantons. Using
eq.~(\ref{Fexpansion2}), one readily concludes that the
contribution of order $h^k$ to the self-duality equation for
$F_{\m\n}[a_\s[h]]$ reads
\begin{equation}
    f^{(k)}_{\m\n}+\sum_{m=1}^k F^{(m,k-m)}_{\m\n}=
\tilde{f}^{(k)}_{\m\n}+\sum_{m=1}^k \tilde{F}^{(m,k-m)}_{\m\n}.
\label{korder}
\end{equation}
Let us assume that $f^{(0)}_{\m\n}$,..., $f^{(k-1)}_{\m\n}$ are
self-dual; then, the statement made below eq.~(\ref{f1}) leads to
the following result: $F^{(m,k-m)}, 1\leq m\leq k$ are self-dual.
This result and eq.~(\ref{korder}) imply that whatever the value of $k>1$
\begin{displaymath}
f^{(k)}_{\m\n}=\tilde{f}^{(k)}_{\m\n}.
\end{displaymath}
We have thus shown --recall eqs.~(\ref{f0})
and~(\ref{f1})-- that $F_{\mu\n}[a_\s [h]]$ in eq.~(\ref{Fexpansion2}) is
self-dual if, and only if, both $\theta_{\m\n}$ and 
$f^{(k)}_{\m\n}$, $\forall k$, are self-dual. Hence, the noncommutative self-duality equation
boils down to the ordinary self-duality equation, when one looks for solutions
that are formal power series in $h\theta^{\m\n}$  and have a non-vanishing 
Pontrjagin number.

One may readily adapt the procedure carried out above to analyze
the existence of solutions to the noncommutative anti-self-duality
equation that are power series  in $h\theta^{\m\n}$. The noncommutative
anti-self-duality equation reads $F_{\m\n}[a_{\s}[h]]=-
\tilde{F}_{\m\n}[a_{\s}[h]]$, with $F_{\m\n}[a_{\s}[h]]$ defined
as in  eq.~(\ref{fmnexpan}). It turns out that in the case at hand
eq.~(\ref{tet-atet}) is replaced with
\begin{equation*}
\sum_a\frac{1}{2}\big[(f^{(0)\,a}_{12})^2+(f^{(0)\,a}_{13})^2+(f^{(0)\,a}_{23})^2\big](\theta+\tilde{\theta})_{\mu\nu}=0.
\end{equation*}
Hence, only when $\theta_{\m\n}$ is anti-self-dual does the
previous equation hold for $f^{(0)}_{\m\n}\neq0$. By replacing
self-dual objects with anti-self-dual tensors in the analysis
above, one finally reaches, with regard to the noncommutative
anti-self-duality equation , the conclusion stated at the beginning
of this Appendix.

\section{Appendix B}
\renewcommand{\theequation}{B.\arabic{equation}}

In this Appendix we shall show that, in the temporal gauge in Minkowski space-time
--$a_0(t,\vec{x})=0$-- and for commutative time
--$\theta^{0i}=0$--, any gauge field $a_i[h](t,\vec{x})$ that
is a formal power series in $h\theta^{\m\n}$ defines a classical vacuum 
of the noncommutative
$SU(N)$ Yang-Mills theory if, and only if, there exists
$g(\vec{x})\epsilon\,SU(N)$ such that
$a_i[h](t,\vec{x})=ig(\vec{x})\partial_i g(\vec{x})^{\dagger}$.

Let us work out first the Hamiltonian of the theory defined by the
action in eq.~(\ref{Sinst1}) rotated to Minkowski space-time and the standard Seiberg-Witten map
in eq.~(\ref{cswmap}). It can be shown that if $\theta^{0i}=0$ 
and $a_0(t,\vec{x})=0$, no time derivatives occur in $A_i$ and
$A_0$ is linear in $\partial_0a_i\equiv \dot{a}_i$:
\begin{equation*}
A_i=a_i+ M_i[a_k,\partial_k a_j];\quad \quad A_0= L^l_0[a_i,\partial_i a_j,\partial_k]\dot{a}_l.
\label{Aex}
\end{equation*}
The field strength $F_{\mu\nu}$ of this noncommutative gauge field
reads
\begin{equation}
\begin{array}{l}
    F_{ij}=f_{ij}+{R}_{ij}[a_k,\partial_k a_m],\quad R_{ij}=\sum_{l>0}h^l R_{ij}^{(l)},
    \\ F_{0i}=\dot{a}_i+{S}_{ij}^c[a_k,\partial_k a_l,\partial_k] \dot{a}^c_j,\quad\quad S^c_{ij}=\sum_{l>0}h^lS_{ij}^{c\,(l)}.
\end{array}
    \label{Fgauge}
\end{equation}
Substituting this expression in eq.~(\ref{Sinst1}) rotated to Minkowski space-time, one obtains
the Lagrangian ${\cal L}[t,\vec{y}]$ of the theory in the temporal
gauge. This Lagrangian  is quadratic in $\dot{a}_i$, so that the
conjugate momenta of the field variable $a_i$ are defined as
usual:
\begin{equation*}
\begin{array}{l}
\pi^a_i(t,\vec{x})=\frac{\delta}{\delta\dot{a}^a_i}\idy{\cal L}[t,\vec{y}]=-\frac{2}{g^2}
\Tr\{F^{0i}(t,\vec{x}) T^a+\idy{S_{ij}}^a[a(y),\partial^y a,\partial^y]\delta^3(\vec{x}-\vec{y})F^{0j}(t,\vec{y})\},\\
\idx
\pi^a_i(t,\vec{x})\dot{a}^a_i(t,\vec{x})=-\frac{2}{g^2}\Tr\idy
F_{0j}F^{0j}(t,\vec{y}).
\end{array}
\end{equation*}
We then define the hamiltonian, ${\cal H}$, of the theory as
follows
\begin{equation}
{\cal H}=\idx \pi^a_i(t,\vec{x})\dot{a}^a_i(t,\vec{x})-\idx{\cal
L}(t,\vec{x})=
\frac{1}{g^2}\idx\Tr\Big(F_{0i}F_{0i}+\sum_{i<j}F_{ij}F_{ij}\Big).
\label{H}
\end{equation}
Notice that ${\cal H}$ is gauge invariant --recall that
$\theta^{0i}=0$-- and that it is equal to
$\int\,d^3\vec{x}\,\Tr T^{00}(t,\vec{x})$, where  $T^{00}(t,\vec{x})$
is the $00$ component of the gauge covariant energy-momentum tensor --see~\cite{Abou-Zeid:2001up} for further discussion--:
$
T^{\mu\nu}=-\frac{1}{g^2}\Big(F^{\mu\alpha}\star
{F^\nu}_\alpha+F^{\nu\alpha}\star
{F^\mu}_\alpha-\frac{1}{2}\eta^{\mu\nu}F^{\a\b}\star F_{\a\b}\Big).
$\par
The classical vacua of the theory are defined by those
$a_i[t,\vec{x}]$ that minimize the hamiltonian given in
eq.~(\ref{H}). Since the fields $F_{\mu\nu}$ are self-adjoint,
${\cal H}$ is positive-definite. Hence, the vacuum configurations
are those which verify
\begin{equation*}
F_{0i}(t,\vec{x})=0,\quad\quad\quad F_{ij}(t,\vec{x})=0.
\end{equation*}

Let us assume that $a_i$ and $f_{\m\n}$ are given by the
expansions in non-negative powers of $h\theta^{\m\n}$ in
eqs.~(\ref{expasionofa}) and~(\ref{fmnexpan}). Then,
eq.~(\ref{Fgauge}) leads to the following expansions:
\begin{equation*}
\begin{array}{l}
F_{0i}=\dot{a}^{(0)}_i+\sum_{l>0}h^l\dot{a}^{(l)}_i+\sum_{l>0}\sum_{s\geq 0}\sum_{t\geq0}h^{l+s+t}S_{ij}^{c\,(l,s)}\dot{a}^{c\,(t)}_j\\
=\dot{a}^{(0)}_i+\sum_{l>0}h^l\Big(\dot{a}^{(l)}_i+\sum_{s=1}^{l}\sum_{t=0}^{l-s}S_{ij}^{c\,(s,t)}\dot{a}_j^{c\,(l-s-t)}\Big),\\
S^{c\,(l,k)}_{ij}=\frac{1}{k!}\frac{d^k}{dh^k}S^{c\,(l)}_{ij}[a^{(0)}+h^l
a^{(l)}]\vert_{h=0}.
\end{array}
\end{equation*}
Hence,  $F_{0i}=0$ is equivalent to the following set of
equalities:
\begin{equation*}
\begin{array}{l}
\dot{a}_i^{(0)}=0,\\
\dot{a}^{(l)}_i+\sum_{s=1}^{l}\sum_{t=0}^{l-s}S_{ij}^{c\,(s,t)}\dot{a}_j^{c\,(l-s-t)}=0,\quad l \geq1.
\end{array}
\end{equation*}
It is easy to show by induction that the solution to the previous
collection of equations reads
$$\dot{a}_i^{(l)}=0\,\,\,\Rightarrow\,\,\,a_i^{(l)}=a_i^{(l)}(\vec{x}),\,\,\,l\geq0.$$
This leads to the conclusion that, in the temporal gauge, the
classical vacua of our noncommutative theory are given by time
independent gauge fields, say, $a_i[h](\vec{x})$, at least if they
can be formally expanded as in eq.~(\ref{expasionofa}).

Now, for a field configuration of the form  of eq.~(\ref{expasionofa})
with field strength as in eq.~(\ref{fmnexpan}), $F_{ij}$ in
eq.~(\ref{Fgauge}) takes the form:
\begin{equation*}
\begin{array}{l}
F_{ij}=f^{(0)}_{ij}+\sum_{l>0}h^l f^{(l)}_{ij}+\sum_{l>0,k\geq0}
h^{l+k}F^{(l,k)}_{ij}=
f^{(0)}_{ij}+\sum_{l>0}h^l\Big(f^{(l)}_{ij}+\sum_{k=1}^l F_{ij}^{(k,l-k)}\Big),\\
F^{(l,k)}_{ij}=\frac{1}{k!}\frac{d^k}{dh^k}F_{ij}^{(l)}[a^{(0)}_i+h^l
a^{(l)}_i]\vert_{h=0},\;F_{ij}^{(l)}[a_i]=\frac{1}{l!}\frac{d^{l-1}}{dh^{l-1}}\big[\frac{dF_{ij}}{dh}\big][a^{(0)}_i+t^l
a^{(l)}_i]\vert_{h=0,t\rightarrow h}
\end{array}
\end{equation*}
where $\frac{dF_{\mu\nu}}{dh}$ is given in
eq.~(\ref{firstderiv}). We thus conclude that $F_{ij}=0$ is
equivalent to the set of equations
\begin{equation}
\begin{array}{l}
    f^{(0)}_{\mu\nu}=0,\\
    f^{(l)}_{ij}+\sum_{k=1}^l F_{ij}^{(k,l-k)}=0,\quad l\geq 1.
\end{array}
\label{setofeq}
\end{equation}

Now, it can be shown by induction that $f_{\m\n}=0$ implies
$F_{\m\n}^{(l)}[a_\r]=0$,
$\forall l$. Using this result, one can prove that
\begin{equation*}
F^{(l,k)}_{ij}=0,\quad \text{if}\quad f^{(n)}_{ij}=0,\quad 0\leq
n\leq k.
\end{equation*}
    Furnished with this result, one readily shows that the solution, $a_i(\vec{x})$, to eq.~(\ref{setofeq})
    must satisfy
\begin{equation*}
     f_{ij}^{(l)}=0\quad\forall l\geq 0\quad \Leftrightarrow\quad
     f_{ij}=0,
\end{equation*}
 i.e., $a_i(\vec{x})$ is a pure gauge:$a_i(\vec{x})=i g(\vec{x})\partial_i
 g^{\dagger}(\vec{x})$.

\newpage

\section{Appendix C}
\renewcommand{\theequation}{C.\arabic{equation}}

In this Appendix, we shall work out the zero modes --that we shall
call bosonic zero modes-- of the operator ${\cal
M}^{ab}_{\m\n}[a_{\s}^{({\rm gsing})}]$ in eq.~(\ref{quadoper})
and the zero mode --referred to as the fermionic zero mode-- of
the operator ${\cal K}[a_{\m}^{({\rm gsing})}]$ defined in
eq.~(\ref{defop}).

\subsection{Bosonic zero modes}

As in the ordinary case --see ref.~\cite{Bernard:1979qt}--, the zero
modes, $\delta^i a_\mu(x)$, of the operator ${\cal
M}^{ab}_{\m\n}[a_{\s}^{({\rm gsing})}]$ in eq.~(\ref{quadoper})
are given by
\begin{equation*}
\delta^i a_\mu=\frac{\partial a^{({\rm
gsing})}_\mu(x,\gamma_j)}{\partial \gamma_i}-\Dcal[a_{\s}^{({\rm
gsing})}]_\mu\Omega^{i}.
\end{equation*}
$\{\gamma_i\}$ denote the collective coordinates of $a_{\m}^{({\rm
gsing})}$ and $\Omega^{i}$ defines a gauge transformation that makes
$\delta^i a_\mu$ satisfy the background field gauge condition
$\Dcal[a_{\s}^{({\rm gsing})}]_{\m}\delta^i a_\mu=0$.

There are, of course, twelve zero modes --as many as the number of
dimensions of the moduli space of our
first-order-in-$\theta$-deformed instanton. Let $U$ denote the
rigid $SU(3)$ transformation that relates $a^{({\rm gsing})}_\nu$
and $a^{({\rm sing})}_\nu$ --see eq.~(\ref{singinst})-- and let
$a^{\bp}_\nu$ denote the upper-left-hand corner embedding in
$SU(3)$ of the ordinary BPST instanton in the singular gauge.
Then, the  zero modes we look for read
\begin{equation*}
\begin{array}{l}
{\delta^{\mu} a_\nu=\frac{\partial a^{({\rm gsing})}_\nu}{\partial
{x_0}_\mu}+U\big(\Dcal[a^{\bp}_\s]_\nu a
^{\bp}_\mu\big) U^{\dagger},}\\
{\delta^{\rho} a_\nu=\frac{\partial a^{({\rm
gsing})}_\nu}{\partial \rho},}\\
{\delta^{a} a_\mu=U\big(\Dcal[a^{\bp}_\s ]_\mu\big[\frac{2r^2}{r^2+\rho^2}\, T^a]\big]\big)U^{\dagger},\quad a\in\{1,2,3\},}\\
{\delta^{\alpha} a_\mu=U\big(\Dcal[a^{(sing)}_\s
]_\mu\big[2\sqrt{\frac{r^2}{r^2+\rho^2}}\,T^\alpha+\frac{ih}{4\sqrt{3}\rho^2}
\frac{(4\rho^2 r^3+3r^5)}{(r^2+\rho^2)^{5/2}}\,(\theta-\tilde{\theta})_{\rho\sigma}\bar{\eta}_{a\rho\sigma} f^{a \alpha c}f^{8cd} T^d\big]\big) U^{\dagger}}\\
{\alpha\in\{4,5,6,7\},\,\, a,c,d\in\{1,2,3\}.}\\
\end{array}
\end{equation*}
We have used the following notation $r\equiv\sqrt{(x-x_0)^2}$. $T^a$, $a=1,2,3$ are given by the embedding of the  generators of
$SU(2)$ into the upper-left-hand corner of the generators of
$SU(3)$ in the fundamental representation. The generators $T^4$
and $T^5$, and $T^6$ and $T^7$, form doublets under the action of
the $SU(2)$ subgroup generated by $T^a$, $a=1,2,3$. The remaining
generator $T^8$ is a singlet under the action of this $SU(2)$
subgroup. The limit $h\theta^{\m\n}\rightarrow 0$ of the zero modes
above yields the ordinary zero modes computed in ref.~\cite{Bernard:1979qt}.

The Jacobian, $J(\gamma)$, of the collective coordinates is given
by the following expression
\begin{equation*}
J(\gamma)=\prod_i\frac{1}{\sqrt{2\pi}}\sqrt{\det
\mathfrak{g}(\gamma_k)},
\end{equation*}
 where the metric, $\mathfrak{g}(\gamma_k)$, of the moduli space is given by:
\begin{equation*}
\mathfrak{g}^{ij}(\gamma_k)=\frac{2}{g^2}\,\Tr \idx
\delta^{i}a_\nu(x,\gamma_k)\delta^{j}a_\nu(x,\gamma_k).
\end{equation*}
    Of course, there are no contributions to $\mathfrak{g}(\gamma_k)$ nor  $J(\gamma)$ that are linear in $h\theta^{\m\n}$:
    they are proportional to $h\theta_{\m\n}\,g_{\m\n}$,
    $g_{\m\n}$ being the space-time metric. Hence, 
\begin{equation*}
J(\gamma)=\frac{1}{(2\pi)^6}\sqrt{\det
    g}=\frac{2^{14}\pi^6\rho^3}{g^{12}}\big[128\,\r^4-25h^2\,(\theta_{\m\n}-
\tilde{\theta}_{\m\n})^2\big]=
\frac{2^{14}\pi^6\rho^7}{g^{12}}+O(h^2\theta^2).
\end{equation*}

\subsection{Fermionic zero mode}

It was shown in ref.~\cite{Martin:2005jy} that, as a consequence of
there being an $U(1)_A$ anomaly, the index of the operator ${\cal K}[a_{\m}]$
is one if $a_{\m}$ has Pontrjagin number equal to one. Hence, at
least in perturbation theory of $h\theta^{\m\n}$, ${\cal
K}[a_{\m}]$ has a unique zero mode --which turns out to be right handed-- 
with unit norm. In this Appendix we shall explicitly construct such zero
mode at first order in $h\theta^{\m\n}$ when $a_{\m}=a_{\m}^{({\rm
gsing})}$. The unit norm zero mode of ${\cal K}[a_{\m}^{({\rm
gsing})}]$ defined in eq.~(\ref{defop}) can be obtained from the
unit norm zero mode of ${\cal K}[a_{\m}^{({\rm sing})}]$ by
applying an appropriate rigid $SU(3)$ transformation. Let us then
solve ${\cal K}[a_{\m}^{({\rm sing})}]\psi_{0}=0$ at first order
in $h\theta^{\m\n}$. To do so, we shall  expand $\psi_{0}$ in
positive powers of $h\theta^{\m\n}$ with coefficients that are
square integrable functions. Up to first order in
$h\theta^{\m\n}$, we have
\begin{equation*}
\psi_{0}=\psi_0^{(0)}+h\psi_0^{(1)}+O(h^2\theta^2),
\end{equation*}
where $\psi_0^{(1)}$ is linear in $\theta^{\m\n}$, and $\psi_0^{(0)}$ and
$\psi_0^{(1)}$ satisfy the following equations:
\begin{equation}
\begin{array}{l}
{\Dirac[a^{\bp}_\s]\psi_0^{(0)}=0,}\\
{\Dirac[a^{\bp}_\s]\psi_0^{(1)}=i\,\,\g_\mu b_\mu
\psi_0^{(0)}+\frac{1}{2}\theta_{\alpha\beta}\g_\mu
f_{\mu\alpha}^{\bp}D[a^{\bp}_\s]_\beta\psi_0^{(0)}-\frac{1}{8}\theta_{\a\b}\g_\mu(\Dcal[a^{\bp}_\s]_\mu
f^{\bp}_{\a\b})\psi_0^{(0)}.}\\
\end{array}
\label{modocero}
\end{equation}
$f^{\bp}_{\a\b}$  denotes the field strength of $a^{\bp}_{\mu}$, both being in the singular gauge.
Recall that $a_{\m}^{({\rm sing})}=a^{\bp}_{\mu}+h\,b_{\mu}$, with
\begin{equation*}
a^{\bp}_{\mu}(x)=\frac{\overline{\eta}_{a\mu\nu}\,(x-x_0)_\nu \r^2}{(x-x_0)^2[(x-x_0)^2+\rho^2]}\,\tau^a,\,\,
b_{\mu}(x)=\frac{2}{\sqrt{3}}\,
(\theta-\tilde{\theta})_{\mu\a}(x-x_0)_\a\frac{(x-x_0)^2+3\rho^2}{((x-x_0)^2+\rho^2)^3}\,T^8.
\end{equation*}

    The first equality in eq.~(\ref{modocero}) is the ordinary zero mode equation. Its solution is well
    known~\cite{'tHooft:1976fv}; using the conventions in eq.~(\ref{alphas}) in Appendix E, it is given by the following spinor with positive
    chirality:
\begin{equation}
    \psi_{0,i m}^{(0)}=\frac{\rho}{\pi r(r^2+\rho^2)^{3/2}}\left[\left(\frac{1+\g_5}{2
    }\right)(\xslash-\xslash_0)\right]_{ij}\epsilon_{j m}.
    \label{thooftmode}
\end{equation}
Note that $i,j$ stand for spinor indices and $m,n$ for colour
indices.

Now, using the properties of the Gell-Mann matrices, one may show
that the second equality in eq.~(\ref{modocero}) implies that the
third colour component of $\psi^{(1)}_0$ must vanish, if it
vanishes at infinity. Hence, the second equality in
eq.~(\ref{modocero}) can be reduced to an equation with colour
indices belonging to $SU(2)$ --in the fundamental representation--
upon replacing $T^8$ in $b_{\m}$ with $\frac{1}{2\sqrt{3}}\,\id$.
This we shall do.

The term on the far r.h.s of eq.~(\ref{modocero}) can be expressed
as follows
$\Dirac^{\bp}\left(-\frac{1}{8}\,\theta_{\alpha\beta}f_{\alpha\beta}\psi^{(0)}_0\right)$.
Let  $\psi^{(1b)}$ be defined by the following equations:
\begin{equation}
\psi_0^{(1)}\equiv \psi^{(1a)}_0+\psi^{(1b)}_0; \quad
\psi^{(1a)}=-\frac{1}{8}\,\theta_{\alpha\beta}f_{\alpha\beta}\psi_0^{(0)}.
\label{simplification}
\end{equation}
In terms of $\psi^{(1b)}_0$, eq.~(\ref{modocero}) reads
\begin{equation}
 \Dirac[a^{\bp}_\s]{\psi^{(1b)}_0}=i\,\g_\mu b_\mu
\psi_0^{(0)}+\frac{1}{2}\theta_{\alpha\beta}\g_\mu
f_{\mu\alpha}^{\bp}D[a^{\bp}_\s]_\beta \psi_0^{(0)}\equiv {\mathfrak R}.
\label{R}
\end{equation}

To solve the previous equation we shall adapt to our case the
technique developed  in ref.~\cite{Grossman:1977mm}. Let us
decompose first $\psi_0^{(0)}$, $\mathfrak{R}$ and $\psi^{(1b)}_0$ into their positive, $R$, and negative, $L$, chirality parts:
\begin{equation*}
\psi_0^{(0)}\equiv\left[\begin{array}{c}
0\\
\psi_{0,R}^{(0)}
\end{array}\right],\,
\mathfrak{R}\equiv\left[\begin{array}{c}
\mathfrak{R}_{L}\\
0\end{array}\right],\,
{\psi^{(1b)}_0}\equiv\left[\begin{array}{c}
\psi^{1b}_{L}\\
\psi^{1b}_{R}
\end{array}\right].
\end{equation*}\par
   Then, when expressed in terms of the bi-spinors $\psi'_{L/R}$ and $\mathfrak{R}'_{L}$,
   defined by the equations
\begin{equation}
\begin{array}{l}
(\psi'_{L/R})_{i m}\equiv (\psi^{1b}_{L/R})_{j m}(\sigma^2)_{j i}\\
 (\mathfrak{R}'_{L})_{i m}\equiv (\mathfrak{R}_{L})_{j m}(\sigma^2)_{ji},
 \end{array},
 \label{phiRp}
\end{equation}
eq.~(\ref{R}) yields
\begin{equation}
\begin{array}{l}
{(\partial_\mu-i a^{\bp}_\mu){\psi'_L}{\alpha}_\mu=0},\\
{(\partial_\mu-i a^{\bp}_\mu){\psi'}_R
\overline{\alpha}_\mu=\mathfrak{R}'_L.}
\end{array}
\label{bispq}
\end{equation}
where $\alpha_\m$ is defined in  Appendix E.

It is well known~\cite{Coleman:1978ae}   that there is no
non-vanishing square integrable ${\psi'}_L$ that solves
the first equation in eq.~(\ref{bispq}). To find ${\psi'}_R$ that
verifies the second equality in eq.~(\ref{bispq}), let us first
express ${\psi'}_R$,  $\mathfrak{R}'_L$ and $a_\mu^{\bp}$in terms of
$\alpha_\mu,\,\overline{\alpha}_\mu$ --see Appendix E--,
$\sigma_\mu=(\overrightarrow{\sigma},i)$ and
${\sigma}_{\mu\nu}$, respectively:
\begin{equation}
\begin{array}{l}
    ({\psi'}_R)_{i m}=M_\m (\alpha_\m)_{m i};\quad (\mathfrak{R}'_L)_{i m}=k_\mu (\overline{\alpha}_\mu)_{m i}= k_4\,\delta_{i m}+ik_{\mu\nu}\,({\sigma}_{\mu\nu})_{m i}\,/\,k_{\mu\nu}=\frac{1}{2}\overline{\eta}_{b\mu\nu}k_b,\\
    a_\mu^{\bp}=-{\sigma}_{\mu\nu}\varphi_\nu,\,\,;\,\, \varphi_\nu=\partial_\nu \log \lambda,\,\,;\,\, \lambda=1+\frac{\rho^2}{(x-x_0)^2}.
\end{array}
\label{descomp}
\end{equation}\par
    By substituting the previous expressions in eq.~(\ref{bispq}) and then using the equalities in
    eq.~(\ref{alphas}), one obtains the following identity:
    \begin{equation}
    i{\sigma}_{\mu\nu}\left(2\partial_\nu M_\mu-\varphi_\nu M_\mu-k_{\mu\nu}\right)+\left(\partial_\mu M_\mu+\frac{3}{2}\varphi_\nu
    M_\nu-k_4\right)=0.
    \label{modocero1}
\end{equation}
Taking traces leads to the conclusion that each summand in the
previous equation must vanish independently. The following
definitions
\begin{equation}
M_\mu\equiv \lambda^{1/2}N_\mu,\quad N_\mu\equiv\partial_\mu
\frac{\phi}{\lambda}+\partial_\nu X_{\nu\mu},\quad
X_{\nu\mu}\,\text{ anti self-dual},
 \label{N}
\end{equation}
allow us to show that  eq.~(\ref{modocero1}) is equivalent to the
following pair of equalities:
\begin{equation}
\begin{array}{l}
    \square X_{\mu\nu}=-\frac{2 k_{\mu\nu}}{\lambda^{1/2}},\\
    \square \phi=\frac{k_4}{\lambda^{5/2}}-\frac{\partial_\mu(\lambda^2 \partial_\nu
    X_{\nu\mu})}{\lambda},
\end{array}
\label{modocero2}
\end{equation}\par
 where
\begin{equation*}
\begin{array}{l}
    k_4=0,\\
    k_{\rho\sigma}=m_{\rho\sigma}-\frac{1}{2}\,\epsilon_{\rho\s\a\b}m_{\a\b},\\
     m_{\rho\sigma}= -\frac{\sqrt{2}}{2\pi r(r^2+\rho^2)^{9/2}}\left[
    \lambda^3(\theta_{\rho\sigma}r^2-4\Delta x_\beta \Delta x_\rho\theta_{\sigma\beta}+4\Delta x_\beta
    \Delta x_\sigma\theta_{\rho\beta})+\frac{1}{6}\
    \lambda\theta_{\rho\sigma}r^2(r^2+3\lambda^2)\right],\\
    \Delta x_\s\equiv(x-x_0)_\s.
\end{array}
\end{equation*}\par

The general solution to eq.~(\ref{modocero2}) is the sum of a
particular solution and the general solution of the corresponding
homogeneous set of equations. A solution to the first equality in
question is
\begin{equation*}
X_{\mu\nu}(x)_{part}=\frac{1}{2\pi}\,\idy\,\frac{1}{(x-y)^2}\,\frac{k_{\m\n}(y)}{\sqrt{\lambda(y)}}.
\end{equation*}
Recall that $-\frac{1}{4\pi}\frac{1}{(x-y)^2}$ is a Green function
of the Laplace operator in four dimensions. By substituting
$k_4=0$ and the previous value of $X_{\mu\nu}(x)$ in the r.h.s. of
the second equation in eq.~(\ref{modocero2}), one shows that this
r.h.s. vanishes. Hence, $\phi_{part}(x)=0$ and
$X_{\mu\nu}(x)_{part}$ above constitute  a particular solution to
eq.~(\ref{modocero2}).

Adding to this particular solution an appropriate solution to the
corresponding homogeneous set of equations is equivalent to adding
$c(\theta)\,\psi^{(0)}_0(x)$ to the $ \psi_{0}^{(1b)}$ --call it
$ \psi_{0\, part}^{(1b)}$-- constructed from $\phi_{part}(x)=0$
and $X_{\mu\nu}(x)_{part}$ right above by using eqs.~(\ref{N}),
(\ref{descomp}), (\ref{phiRp}) and (\ref{R}). $c(\theta)$ is an
arbitrary coefficient linear in $\theta^{\m\n}$. Normalization to
1 of $\psi_{0}$ renders $c(\theta)$ equal to zero, for
$\psi^{(0)}_0$ has unit norm and $\psi^{(1)}_{0\, part}=\psi^{(1a)}_{0}+\psi^{(1b)}_{0\, part}$ is
orthogonal to $\psi^{(0)}_0$ --$(\psi^{(0)}_0,\psi^{(1)}_{0\,
part})$ is proportional to $\theta_{\m\n}\,g_{\m\n}$, $g_{\m\n}$
being the space-time metric--. We thus conclude that
$\psi_{0}^{(1b)}$ in eq.~(\ref{simplification}) is equal to
$\psi_{0\, part}^{(1b)}$, so that we finally have
\begin{equation}
\begin{array}{l}
\psi_{0,im}^{(1)}=\psi_{0,im}^{(1a)}+\psi_{0,im}^{(1b)}=\\
    \frac{1}{\pi\,\rho\,r\,(r^2+\rho^2)^{7/2}}\left[\left(\frac{1+\g_5}{2
    }\right)\g_\sigma\right]_{ij}\left\{\frac{i}{12} (x-x_0)_\rho \left[(\theta-\tilde{\theta})_{\rho\sigma}
    (4r^4+14r^2\rho^2)-6\rho^4(\theta+\tilde{\theta})_{\rho\sigma}\right]\epsilon_{j m}\right.\\
    \phantom{ \psi_{0,\alpha m}^{(1)}=\frac{1}{\pi\,\rho\,r\,(r^2+\rho^2)^{7/2}}\left[\left(\frac{1+\g_5}{2
    }\right)\g_\sigma\right]_{\alpha\beta}}
    \left.+2\,\rho^4\, \theta_{\a\g}(x-x_0)_\sigma\left[\frac{x_\alpha x_\nu}{r^2}-\frac{1}{4}\,\delta_{\alpha\nu}\right]
    \epsilon_{j n}\,{\tau}_{\g\nu,m n}\right\}.
\end{array}
\label{correctionmode}
\end{equation}
${\tau}_{\mu\nu}$ are the analogs of
${\sigma}_{\mu\nu}$ in $SU(2)$ colour space.

Finally, by acting with the appropriate $SU(3)$ transformation,
$U$, on $\psi_{0}^{(0)}$ and $\psi_{0}^{(1)}$ in
eqs.~(\ref{thooftmode}) and~(\ref{correctionmode}), one obtains
the unit norm zero mode of ${\cal K}[a_{\m}^{({\rm gsing})}]$.
This zero mode reads
\begin{equation}
    \psi_0\equiv\psi^0+\psi^{1a}+\psi^{1b},
    \label{modocerolef}
\end{equation}
where, writing these right handed spinors in two-component notation, we have
\begin{equation}
\begin{array}{l}
    {\psi^0}_{i m}=\phi(x)\,(x-x_0)_\mu ( \overline{\alpha}_\mu)_{i j}\epsilon_{j n}U_{mn},\\
    {\psi^{1a}}_{i m}=h\left[\Gamma_{\alpha\gamma}(x)\,(x-x_0)_\a (x-x_0)_\nu+\Lambda _{\nu\gamma}(x)\right]\,
    (x-x_0)_\mu( \overline{\alpha}_\mu)_{i j}\epsilon_{j o} ({\tau}_{\g\nu})_{no}U_{mn},\\
    {\psi^{1b}}_{i m}= h\chi_{\alpha\sigma}(x)\,(x-x_0)_\alpha(\overline{\alpha}_\sigma)_{ij}\epsilon_{j n}U_{mn}.\\
\end{array}
\label{modocerocomp}
\end{equation}
The functions $\phi(x)$, $\Gamma_{\a\g}(x)$, $\Lambda_{\a\g}(x)$
and $\chi_{\alpha\sigma}(x)$ are defined thus:
\begin{equation}
\begin{array}{l}
    \phi(x)=\frac{\rho}{\pi r(r^2+\rho^2)^{3/2}},\quad
    \Gamma_{\a\g}(x)=\frac{2 \theta_{\a\g}\rho^3}{\pi r^3(r^2+\rho^2)^{7/2}},\quad \Lambda_{\a\g}(x)=
    -\frac{ \theta_{\a\g}\rho^3}{2\pi r(r^2+\rho^2)^{7/2}},\\
    \chi_{\alpha\sigma}(x)=\frac{i}{12\pi\rho r(r^2+\rho^2)^{7/2}}\left[(\theta-\tilde{\theta})_{\a\s}(4r^4+14r^2\rho^2)
    -6\rho^4(\theta+\tilde{\theta})_{\a\s}\right].
\end{array}
\label{modocerofunc}
\end{equation}
Recall that $r=\sqrt{(x-x_0)^2}$.

\section{Appendix D}
\renewcommand{\theequation}{D.\arabic{equation}}

In this Appendix we shall show
that, if $\theta^{4i}=0$, $i=1,2,3$, no topologically nontrivial
field configurations can be found that {\bf a)} are formal power series  in
$h\theta^{\mu\nu}$ and {\bf b)} at second order in $h\theta^{\m\n}$ solve the equations of motion of noncommutative $SU(3)$ Yang-Mills theory.
To show it, we shall use the technique devised in 
ref.~\cite{Derrick:1964ww}. Thus,  we shall consider the
behaviour of the action in eq.~(\ref{Sinst1}) under the following infinitesimal changes of scale:
 \begin{equation}
    a'_\mu=\lambda a_\mu(\lambda x),\quad\lambda=1+\delta\lambda,
\label{da}
\end{equation}
where $a_\mu$ satisfies the equations of motion.
    The action can be written as:
\begin{equation}
S_{NCYM}=\sum_{n=0}^\infty h^n S^{(n)}=\sum_{n=0}^\infty h^n\idx{\cal L}^{(n)}[a,\partial],
\label{esen}
\end{equation}
where ${\cal L}^{(n)}$ is the term of the lagrangian of order $h^n$, which, due to the fact that  ${\cal L}^{(n)}$ contains $n$ powers of $\theta^{\m\n}$, is a polynomial of mass degree $4+2n$ in $a_\mu$ and $\partial_\mu$. Hence, 
\begin{equation*}
\idx{\cal L}^{(n)}[\lambda a(\lambda x),\partial^x]=\lambda^{-4}\idy {\cal L}^{(n)}[\lambda a(y),\lambda\partial^y]=\lambda^{2n}\idy{\cal L}[a(y),\partial^y].
\end{equation*}
Thus, under the change in eq.~(\ref{da}),$S^{(n)}[a]$ in eq.~(\ref{esen}) changes as follows
\begin{equation*}
    S^{(n)'}[a']=\lambda^{2n}S^{(n)}[a].
\end{equation*}
Since we are assuming that the original field configuration $a_\mu$ in eq.~(\ref{da}) is a solution to the equations
of motion,  the following equivalent equations hold
\begin{equation}
S[a]=S[\lambda a(\lambda x)]+O(\delta\lambda^2)\Leftrightarrow \delta\lambda\sum_{n=1}^\infty2n\, h^n S^{(n)}[a]=O(\delta\lambda^2)\,\,\, \forall 
\lambda\,\,\,\Leftrightarrow\,\,
\sum_{n=1}^\infty2n\,h^n\, S^{(n)}[a]=0.
\label{dS}
\end{equation}
Now, let our solution to the equations of motion, $a_{\m}$, be given by the 
following power series: $a_\mu=\sum_{n=0}^\infty h^n a^{(n)}_\mu$. By substituting this power series in $S^{(n)}$ in eq.~(\ref{esen}), one obtains $S^{(n)}[a]=\sum_{k=0}^\infty h^k S^{(n,k)}[a],\,\,S^{(n,m)}=\frac{1}{m!}\frac{d^m}{dh^m}S^{(n)}[a_\mu]|_{h=0}$. Combining this result with the equality on the far right 
of eq.~(\ref{dS}), one ends up with
\begin{equation*}
0=\sum_{k=1}^{\infty}\,h^k\,\Big(\sum_{m=0}^{k-1} (k-m)S^{(k-m,m)}\Big)\quad
\Leftrightarrow\quad 0=\sum_{m=0}^{k-1} (k-m)S^{(k-m,m)}=0, \forall k\geq 1.
\end{equation*}
    For the action $S_{NCYM}$ to be stationary up to order $h^l$, the previous identities have to be verified for
$k\leq l$. In our case, we want the action to be stationary at order $h^2$,
 so that we should check if the following equalities hold
\begin{equation}
S^{(1,0)}=0,\quad 2S^{(2,0)}+S^{(1,1)}=0.
\label{testh2}
\end{equation}
    From eq.~(\ref{Sinst1}) and eq.~(\ref{swpontri}), one concludes that 
$S_{NCYM}$ for a field configuration with a well defined
topological charge $n$ reads
\begin{equation}
\begin{array}{c}
    S_{NCYM}=
    \frac{1}{2g^2}\,\mathrm{Tr}\,\idx \left[ f_{\mu\nu}\tilde{f}_{\mu\nu}+\frac{1}{2}(F_{\mu\nu}-\tilde{F}_{\mu\nu})^2\right]
    =\frac{8\pi^2n}{g^2}+\frac{1}{4g^2}\Tr\,\idx(F_{\mu\nu}-\tilde{F}_{\mu\nu})^2.
\end{array}
\label{Stop}
\end{equation}
$F_{\mu\nu}$ is given by the Seiberg-Witten  map as a power series:
$F_{\mu\nu}[a_\r]=f_{\mu\nu}+\sum_{l>0}h^lF_{\mu\nu}^{(l)}$. When evaluating 
these terms for the solution $a_\mu=\sum_{n=0}^\infty h^n a^{(n)}_\mu$ 
we get again $F_{\mu\nu}^{(n)}[a_\r]=
\sum_{k=0}^\infty h^k F_{\mu\nu}^{(n,k)}[a_\r],\,\,
F_{\mu\nu}^{(n,m)}[a_\r]=\frac{1}{m!}\frac{d^m}
{dh^m}F_{\mu\nu}^{(n)}[a_\r]|_{h=0}$ and $f_{\mu\nu}=f_{\mu\nu}^{(0)}+
\sum_{k=1}^{\infty}\,h^k\,f_{\mu\nu}^{(k)}$. Hence,  $S^{(1,0)}$, 
$S^{(1,1)}$ and $S^{(2,0)}$ in eqs.~(\ref{esen}) and (\ref{testh2}) 
are given by 
\begin{equation*}
\begin{array}{l}
    S^{(1,0)}=\frac{1}{2g^2}\Tr\idx(f^{(0)}_{\mu\nu}-\tilde{f}^{(0)}_{\mu\nu})(F^{(1,0)}_{\mu\nu}-\tilde{F}^{(1,0)}_{\mu\nu}),\\
    S^{(1,1)}=\frac{1}{2g^2}\Tr\idx\Big[(f^{(1)}_{\mu\nu}-\tilde{f}^{(1)}_{\mu\nu})(F^{(1,0)}_{\mu\nu}-\tilde{F}^{(1,0)}_{\mu\nu})+(f^{(0)}_{\mu\nu}-\tilde{f}^{(0)}_{\mu\nu})(F^{(1,1)}_{\mu\nu}-\tilde{F}^{(1,1)}_{\mu\nu})\Big],\\
    S^{(2,0)}=\frac{1}{2g^2}\Tr\idx(f^{(0)}_{\mu\nu}-\tilde{f}^{(0)}_{\mu\nu})(F^{(2,0)}_{\mu\nu}-\tilde{F}^{(2,0)}_{\mu\nu})+\frac{1}{4g^2}\Tr\idx(F^{(1,0)}_{\mu\nu}-\tilde{F}^{(1,0)}_{\mu\nu})^2.
    \end{array}
\end{equation*}
	
In section 2, we saw that the most general solution to the equations of motion at order $h$ is given by 
\begin{equation}
a_\m=U\big(a_\m^{\bp}+hb^8_\mu T^8+h\sum_{a=1}^{7}b^a_\mu T^a\big)U^{\dagger},
\label{classsol}
\end{equation}
where $b^8_\mu$ is given in eq.~(\ref{fordertheta}), $h\sum_{a=1}^{7}b^a_\mu T^a$ is 
any linear combination --with coefficients linear in $h\theta^{\m\n}$-- of the ordinary 
bosonic zero modes~\cite{Bernard:1979qt} --i.e., the solutions to eq.~(\ref{zeromodeseq})-- and $U$ is a rigid $SU(3)$ transformation.
For a field configuration of this type, we have  $S^{(1,0)}=0$, for 
$f_{\mu\nu}^\bp=\tilde{f}^\bp_{\mu\nu}$. The first condition in eq.~(\ref{testh2}) is 
thus automatically satisfied. Notice that eq.~(\ref{Stop}) tells us that any contribution of order $h^2\theta^2$ to the classical solution in eq.~(\ref{classsol}) yields 
a contribution of order $h^3\theta^3$ to the action $S_{NCYM}$.
We will show next that the second condition in eq.~(\ref{testh2}) is violated by 
the solution in eq.~(\ref{classsol}), so that it is impossible to find a field configuration with non-zero topological charge that makes the action stationary up to order $h^2\theta^2$.
First, $F^{(1,1)}$ and $F^{(2,0)}$ do not contribute neither to $S^{(1,1)}$ nor 
$S^{(2,0)}$. Now, taking a closer look at the structure constants of $SU(3)$, one sees that 
the contribution to $S^{(1,1)}$ of $h\sum_{a=1}^{7}b^a_\mu T^a$ is zero:
$$S^{(1,1)}=\frac{1}{2g^2}\Tr\idx(f_{\mu\nu}^{(1),8}T^8-\tilde{f}^{(1),8}_{\mu\nu}T^8)
(F^{(1,0)}_{\mu\nu}-\tilde{F}^{(1,0)}_{\mu\nu}),\,\,f_{\mu\nu}^{(1),8}=U(\Dm b^8_\nu-\Dn b^8_\mu)U^\dagger.$$
By evaluating $S^{(1,1)}$ and $S^{(2,0)}$ for the field configuration in eq.~(\ref{classsol}), one obtains
\begin{equation*}
\begin{array}{l}
    S^{(1,1)}[a_\mu]=-\frac{8\pi^2}{7g^2\rho^4}(\theta_{\mu\nu}-\tilde{\theta}_{\mu\nu})^2\\
    S^{(2,0)}[a_\mu]=\frac{12\pi^2}{7g^2\rho^4}(\theta_{\mu\nu}-\tilde{\theta}_{\mu\nu})^2
    \end{array}
    \quad 2 S^{(2,0)}+S^{(1,1)}=\frac{16\pi^2}{7g^2\rho^4}(\theta_{\mu\nu}-\tilde{\theta}_{\mu\nu})^2,
\end{equation*}
so that eq.~(\ref{testh2}) is violated and, furthermore, this happens independently of the arbitrary part 
of the solution in eq.~(\ref{classsol}).  Hence, the only way to make $2S^{(2,0)}+S^{(1,1)}$ zero is by 
taking $\rho$ to infinity, which would turn our solution into the trivial one. \par

	This conclusion still holds for the most general Seiberg-Witten map at order $h$, given by 
eq.~(\ref{SWgeneral}). It turns out that the expression for $S^{(1)}$ obtained with this map is 
the same as the one derived with the standard map for arbitrary $a_\mu$ tending to zero at infinity. 
Therefore, the field configuration in eq.~(\ref{classsol}) is the most general classical solution 
at order $h\theta^{\m\n}$ with unit topological charge for an arbitrary Seiberg-Witten map. When checking 
whether the conditions in eq.~(\ref{testh2}) hold, the values of $S^{(1,0)}$ and $S^{(1,1)}$ are unchanged 
since so is $S^{(1)}$. It can also be seen that, for field configurations that are $\theta$-dependent 
deformations of the ordinary instanton, the value of $S^{(2,0)}$ is the same for all the 
Seiberg-Witten maps. Therefore the conditions in eq.~(\ref{testh2}) are always violated, and this 
concludes the proof of the statement made at the beginning of this section.

\section{Appendix E}
\renewcommand{\theequation}{E.\arabic{equation}}

In this Appendix sundry formulae are collected.

\subsection{Spinor and $SU(2)$ matrices}

\begin{equation}
\begin{array}{l}
    \sigma_\mu=(\overrightarrow{\sigma},i),\quad\,\,\overline{\sigma}_\mu=(-\overrightarrow{\sigma},i),\quad\,\,\alpha_\mu=(-i\overrightarrow{\sigma},\id)=-i\sigma^-_\mu,\quad\,\,\overline{\alpha}_\mu=(i\overrightarrow{\sigma},\id)=i\sigma^+_\mu,\\
    \overline{\sigma}_{\mu\nu}=\frac{1}{4i}(\overline{\alpha}_\mu\alpha_\nu-\overline{\alpha}_\nu\alpha_\mu)=\frac{1}{2}\,{\eta}_{a\mu\nu}\sigma_a;\,\,\sigma_{\mu\nu}=\frac{1}{4i}({\alpha}_\mu\overline{\alpha}_\nu-{\alpha}_\nu\overline{\alpha}_\mu)=\frac{1}{2}\,\overline{\eta}_{a\mu\nu}\sigma_a,\\
    \alpha_{\mu}\overline{\alpha}_\nu=g_{\mu\nu}+2i{\sigma}_{\mu\nu},\\
	\text{and analogously for the $SU(2)$ generators $\tau^a, a=1,2,3$.}
\end{array}
\label{alphas}
\end{equation}
\begin{equation*}
\begin{array}{l}
    \epsilon_{12}=+1;\quad\epsilon_{im}\epsilon_{jn}=\frac{1}{8}(\sigma_\mu^- \sigma_\nu^+)_{ij}(\tau_\mu^-\tau_\nu^+)_{mn}.
\end{array}
\end{equation*}
\begin{equation*}
\begin{array}{l}
   \gamma_\mu=\left[\begin{array}{cc}&\alpha_\mu\\\overline{\alpha}_\mu\end{array}\right],\quad\g_5=-\g_1\g_2\g_3\g_4=\left[\begin{array}{cc}-\id&\\&\id\end{array}\right],\quad\g_{\mu\nu}=\frac{1}{4i}[\g_\mu,\g_\nu]
\end{array}
\end{equation*}

\subsection{$SU(3)$ averages}

\begin{equation}
\begin{array}{l}
    \int\!dU =1,\\
    \int\!dU U_{i a} {U^\dagger}_{j b} =\frac{1}{3}\,\delta_{j a}\delta_{i b},\\
    \int\!dU U_{i_1 a_1} {U^\dagger}_{j_1 b_1} U_{i_2 a_2} {U^\dagger}_{j_2 b_2} =\frac{1}{3^2}\,\delta_{j_1 a_1}\delta_{i_1 b_1}\delta_{j_2 a_2}\delta_{i_2 b_2}+
    \frac{1}{4\cdot8}\,(\lambda^a)_{j_1 a_1}(\lambda^a)_{j_2 a_2}(\lambda^b)_{i_1 b_1}(\lambda^b)_{i_2 b_2}\\
    \int\!dU U_{i_1 a_1} {U^\dagger}_{j_1 b_1} U_{i_2 a_2} {U^\dagger}_{j_2 b_2} U_{i_3 a_3}{U^\dagger}_{j_3 b_3}=
    \frac{1}{3^3}\,\delta_{j_1 a_1}\delta_{i_1 b_1}\delta_{j_2 a_2}\delta_{i_2 b_2}\delta_{j_3 a_3}\delta_{i_3 b_3}\\
    \phantom{\int\!dU U_{i_1 a_1} {U^\dagger}_{j_1 b_1} }
    +\frac{1}{4\cdot3\cdot8}\left[(\lambda^a)_{j_1 a_1}(\lambda^a)_{j_2 a_2}(\lambda^b)_{i_1 b_1}(\lambda^b)_{i_2 b_2}
    \delta_{j_3 a_3}\delta_{i_3 b_3}+(3\leftrightarrow 1)+(3\leftrightarrow 2)\right]\\
    \phantom{\int\!dU U_{i_1 a_1} {U^\dagger}_{j_1 b_1} }
    +\frac{3}{8\cdot5\cdot8}\,d_{ijk}d_{abc}(\lambda^i)_{j_1 a_1}(\lambda^j)_{j_2 a_2}(\lambda^k)_{j_3 a_3}(
    \lambda^a)_{i_1 b_1}(\lambda^b)_{i_2 b_2}(\lambda^c)_{i_3 b_3}\\
    \phantom{\int\!dU U_{i_1 a_1} {U^\dagger}_{j_1 b_1} }
    +\frac{1}{8\cdot3\cdot8}\,f_{ijk}f_{abc}(\lambda^i)_{j_1 a_1}(\lambda^j)_{j_2 a_2}(\lambda^k)_{j_3 a_3}
    (\lambda^a)_{i_1 b_1}(\lambda^b)_{i_2 b_2}(\lambda^c)_{i_3
    b_3}.
\end{array}
\label{intgrupo}
\end{equation}


\subsection{Fourier transform and low momenta approximations}

    Fourier transform is defined as follows

\begin{equation*}
    f(p)=\idx e^{ipx}f(x),\quad \quad f(x)=\idp e^{-ipx}f(p).
\end{equation*}\par
    In terms of the modified Bessel functions $I_\nu(z)$ and $K_\nu(z)$, the Fourier transform
    of the functions
    $\phi(x), \,\Gamma(x),\,\Lambda(x),\,\chi(x)$ introduced in
eq.~(\ref{modocerofunc}) read
\begin{equation}
\begin{array}{l}
    \phi(p)=2\pi\rho\left[I_0\left(\frac{u\rho}{2}\right)K_0\left(\frac{u\rho}{2}\right)-I_1\left(\frac{u\rho}{2}\right)K_1\left(\frac{u\rho}{2}\right)\right],\\
    \Gamma_{\a\sigma}(p)=\frac{32 \pi\rho}{15}\,\theta_{\a\sigma}\left(\frac{d}{d\rho^2}\right)^2 I_1\left(\frac{u\rho}{2}\right)K_1\left(\frac{u\rho}{2}\right),\\
    \Lambda_{\a\sigma}(p)=-\frac{16\pi\rho^3}{15u}\,\theta_{\a\sigma}\frac{d}{du}\left(\frac{d}{d\rho^2}\right)^3I_0\left(\frac{u\rho}
    {2}\right)K_0\left(\frac{u\rho}{2}\right),\\
    \chi_{\a\s}(p)=\frac{8i\pi}{45 u\rho}\left\{(\theta-\tilde{\theta})_{\a\s}
    \left[4\left[\left(\frac{d}{d\rho^2}\right)^2+\frac{3}{u}\frac{d}{du}\right]^2-14\rho^2
    \left[\left(\frac{d}{d\rho^2}\right)^2+\frac{3}{u}\frac{d}{du}\right]\right]-6(\theta+\tilde{\theta})_{\a\s}\right\}\times\\
    \phantom{\chi_{\a\s}(p)=}\frac{d}{du}\left(\frac{d}{d\rho^2}\right)^3I_0\left(\frac{u\rho}{2}\right)K_0\left(\frac{u\rho}{2}\right).
\end{array}
\label{fourier}
\end{equation}\par
The variable $u$ stands for $\sqrt{p^2}$.
Let ${}^{'}$ denote derivative with respect to $u$. We have the
following low momenta --$u\r\ll 1$-- expansions:
\begin{equation}
\begin{array}{l}
    \phi'(u)\sim-\frac{2\pi\rho}{u}+O(u),\\
    \Gamma'''_{\a\s}(u)\sim O(u),\,\,\Gamma''(u)\sim O(u^0),\,\,\Gamma'(u)\sim O(u),\\
    \Lambda'_{\a\s}(u)\sim O(u),\\
    \chi'_{\a\s}(u)\sim-\frac{2i\pi}{3\rho
    u}\,(\theta-\tilde{\theta})_{\a\s}+O(u).
\end{array}
\label{asint}
\end{equation}

\end{document}